\documentclass[aps,amsmath,amsfonts,amssymb,groupedaddress,nofootinbib,nobalancelastpage,floatfix,superscriptaddress,secnumarabic]{revtex4}

\usepackage{amstext,amssymb}
\usepackage{amsmath}
\usepackage{graphicx}
\usepackage[hyperfootnotes=true]{hyperref}
\usepackage{xspace}
\usepackage{color}
\usepackage{slashed}
\usepackage{tabularx}
\def\bpm{\begin{pmatrix}}
\def\epm{\end{pmatrix}}
\def\bsp#1\esp{\begin{split}#1\end{split}}
\usepackage{booktabs}
\usepackage{setspace}

\usepackage{ulem}

\begin{document}
\title{ Collider Signatures of $W_R$ boson in the Alternative Left-Right Model}
\author{Mariana Frank}
\email{mariana.frank@concordia.ca}
\affiliation{Department of Physics, Concordia University,
7141 Sherbrooke St. West, Montreal, QC H4B 1R6, Canada}

\author{Chayan Majumdar}
\email{c.majumdar@ucl.ac.uk}
\affiliation{Department of Physics and Astronomy, University College London, London WC1E 6BT, United Kingdom}

\author{Poulose Poulose}
\email{poulose@iitg.ac.in}
\affiliation{Department of Physics, Indian Institute of Technology Guwahati, Assam 781 039, India}

\author{Supriya Senapati}
\email{ssenapati@umass.edu}
\affiliation{Amherst Center for Fundamental Interactions, Department of Physics, University of Massachusetts, Amherst, MA 01003, USA}

\author{Urjit A. Yajnik}
\email{urjit.yajnik@iitgn.ac.in}
\affiliation{Department of Physics, Indian Institute of Technology Gandhinagar, Gujarat, 382055 India}

\begin{abstract}
Alternative Left-Right Models offer an attractive option to left-right models. Emerging from $E_6$ grand unification, these models are consistent with light scalars which do not induce  flavour-changing neutral currents due to the presence of exotic quarks. Here we investigate the signature at the LHC collider of the charged $W_R$ boson, which can be lighter than in left-right models. We include constraints from collider data and show that $W_R$ can be produced in pairs, or in conjunction with a light charged Higgs boson. The final decay products involve leptons or jets.  We explore all production and decay possibilities and indicate which ones are most promising to be observed at the colliders. Our analysis shows that signals of $W_R$ bosons can be observed at the LHC at 27 TeV, some for lower luminosity, and under most favourable conditions, even at 13 TeV.
\end{abstract}

\pacs{}
\maketitle
\section{Introduction}
\label{sec:intro}          
The Standard Model (SM) of particle physics describes the properties of elementary particles of nature and their interactions.
 Still there are some puzzles which cannot be addressed within this SM framework. Being a chiral framework, discrete parity symmetry is broken maximally in low-energy weak phenomenology while the strong and electromagnetic forces are parity conserving. Also due to absence of any right-handed partners of SM left-handed neutrinos in this scenario, SM cannot explain the observational inconsistency posed by various long and short baseline solar, atmospheric, accelerator and reactor neutrino oscillation experiments like SuperK \cite{Super-Kamiokande:2001ljr} and SNO \cite{SNO:2002tuh}, which indicate that active neutrinos have small but finite mass. These puzzles definitely seek the extension of SM beyond the usual framework known as Beyond the SM (BSM) scenarios.

As one of the most phenomenologically attractive theories in BSM perspective, one can extend the SM to the Left-Right Symmetric Model (LRSM) \cite{Pati:1974yy,Mohapatra:1974gc,Senjanovic:1975rk,
Mohapatra:1977mj}. Based on the gauge structure $SU(3)_C \otimes SU(2)_L \otimes SU(2)_R \otimes U(1)_{B-L}$, such framework can solve above discussed inconsistencies within a unified scheme. These models relate the maximal breaking of parity of low-energy weak phenomenology to the origin of small light active neutrino masses. Now one can think of the SM as a low-energy effective field theory of such LRSMs. Unlike SM, right-handed neutrinos are naturally included in this framework which are responsible for the mass generation of light neutrinos via the so-called seesaw mechanism \cite{Mohapatra:1979ia,Schechter:1980gr}. The breaking of this LRSM gauge structure to SM needs extra Higgs scalars in addition to the usual doublet Higgs introduced in SM. This breaking can be done by either (i)  a Higgs bidoublet $\Phi$ and two doublets $\chi_{L,R}$, or (ii) a Higgs bidoublet and two triplets $\Delta_{L,R}$ or (iii) combination of the above two scenarios. The LRSM replaces the hypercharge $Y$ by a gauged $B-L$ gauge group which, unlike the hypercharge, has a  physical significance. In addition, LRSM as a subgroup of $SO(10)$ \cite{Slansky:1981yr} can approximately explain the gauge coupling unification in the context of grand unified scenarios (GUTs). 

In addition to the usual LRSM, one could  envision another model with left-right symmetry, arising from  the breaking of the exceptional group
$E_6$~\cite{Gursey:1975ki,Achiman:1978vg} rather than $SO(10)$. The $E_6$ group has an $SU(3)\otimes SU(3)\otimes SU(3)$ as one of its maximal subgroups. One of these $SU(3)$ groups is unbroken and represents the strong interaction group $SU(3)_c$, while the other two  break into the $SU(2)_L \otimes SU(2)_H \otimes U(1)_X$ group which further contains the $SU(2)_L \otimes U(1)_Y$ electroweak symmetry. In the left-right symmetric model (LRSM),  $SU(2)_H$ is identified with $SU(2)_R$ and $U(1)_X$ with $U(1)_{B-L}$. The right-handed SM fermions and the right-handed neutrino $\nu_R$ belong to $SU(2)_R$ doublets. This framework is quite successful, as it provides a natural explanation for parity and neutrino masses,  but this model  also suffers from unavoidable tree-level flavour-violating interactions (FCNCs) that conflict with the observed
properties of kaon and $B$-meson systems.  The flavour changing neutral currents arise from the fermion-Higgs couplings induced by off-diagonal mixing matrices corresponding to LRSM fermions and bidoublet neutral Higgs.  Consequently, the $SU(2)_R \otimes
U(1)_{B-L}$ symmetry has to be broken at a very high energy scale, which pushes the masses of the extra Higgs and gauge bosons of the model to be very heavy, making them unlikely to be detected at the LHC. In addition the minimal particle spectrum of LRSM lacks a natural DM candidate~\cite{Bahrami:2016has}.

If one identifies the $SU(2)_H$ symmetry with a $SU(2)^\prime_{R}$ group, where the assignments of the SM fermions are different than usual LRSM~\cite{Ma:1986we,Ashry:2013loa,Frank:2005rb}, this problem can be circumvented, but at the cost of introducing new exotic fermions. This model is called the Alternative Left-Right Model (ALRM)~\cite{Babu:1987kp,Ma:2010us}. This model has received some more detailed examination \cite{Ashrythesis,Ashry:2013loa,Frank:2019nid}, and some attractive features were explored. Its vacuum stability has been analysed in \cite{Frank:2021ekj}. The model has been shown to allow for the introduction of an $R$ parity symmetry, similar to the one in supersymmetry, and thus provide  both bosonic and fermionic dark matter candidates \cite{Frank:2022tbm},  whose collider signatures were explored. Also, it has been shown that the model provides significant contributions to the $0\nu\beta\beta$ decay through new $W\,H$ mediation, and  CP violation arising from the right-handed neutrino decay generates resonant leptogenesis, sufficient to yield the correct baryon asymmetry of the universe (BAU) \cite{Frank:2020odd}.

In this work we continue our examination of the ALRM and investigate the collider signatures of the right-handed charged boson $W_R$. First, this is interesting because, while many extended gauge structures provide additional neutral gauge bosons, the presence of a charged gauge boson is indicative of a gauge structure involving $SU(n)$, with $n>2$, limiting the BSM scenarios responsible. In addition, while in LRSM the $W_R$ must be heavy (from the constraints of both collider searches \cite{ATLAS:2019isd,CMS:2021dzb} and FCNCs \cite{ParticleDataGroup:2022pth}), in ALRM $W_R$ couples to exotic fermions and can be light. Its mass is restricted by searches for $Z^\prime $ bosons \cite{CMS:2018ipm,ATLAS:2017fih}, but only loosely, as the mass ratio $M_{W_R}/M_{Z^\prime}$ depends on gauge coupling constants as well. We show here that the $W_R$ in ALRM could be detected through its production followed by decays into either quarks/exotic quarks and leptons/scotinos (new neutral leptons), as it could yield signals at FCC-hh (HE-LHC) \cite{FCC:2018byv} running at $\sqrt{s}=27$ TeV and, in favourable circumstances, even at the HL-LHC. For this analysis, in the present work, we classify all production channels viz., in pairs, or associated with a charged Higgs, and corresponding decays modes, simulate them at the collider, restricting ourselves to the partonic level, and search for the most promising outcome.

The structure of the paper is as follows. We describe our model in Sec. \ref{sec:model} and proceed to analyze the collider implications of the model in Sec. \ref{sec:collider}. We distinguish three different cases, one where $W_R$ is heavier than the exotic quarks and scotinos viz., \ref{subsec:1}, one where the $W_R$ is lighter than the exotic quarks but heavier than scotinos, but with scotinos lighter than the charged Higgs viz., $H_2^\pm$ \ref{subsec:2},  in both of which cases the $W_R$ decay predominantly into leptons; and finally, in \ref{subsec:3}, the case where $W_R$ and $H_2^\pm$ both are lighter than the exotic quarks and scotinos.
As in the other cases, $H_2^\pm$ is lighter than $W_R$. $H_2^\pm$ decays into lighter neutral CP-even or CP-odd Higgs $H_1^0 / A_1^0$, which can be considered as scalar dark matter, stabilised by $R$-parity, in our framework. We summarize our findings and conclude in Sec. \ref{sec:conclusion}.

\section{The Model}
\label{sec:model}

In 1987 Earnest Ma proposed a unified framework $E_6$ to explain the low-scale symmetry in the context of heterotic string theories \cite{Ma:1986we}. This $E_6$ has two maximal subgroup structures $SO(10) \otimes U(1)$, and $SU(3) \otimes SU(3) \otimes SU(3)$. Usual or canonical LRSM can be embedded in both of these subgroups. In the former case, the $\textbf{16}$ representation of $SO(10)$ subgroup contains of the whole fermion structure of the theory. In the latter $SU(3) \otimes SU(3) \otimes SU(3)$ breaking chain, the \textbf{27} representation under $E_6$ is the one where the usual LRSM fermion contents are rearranged \cite{Frank:2005rb}.

However, within this symmetry breaking chain, FCNC constrains the left-right symmetry breaking scale of LRSM to be very high so that any residual signatures of broken $SU(2)_R$ are absent in currently accessible collider searches. Considering instead the symmetry breaking to the gauge group $SU(3)_C \otimes SU(2)_L \otimes SU(2)^\prime_R \otimes U(1)_{B-L} \otimes S$ (ALRM) where we have used primed notation to distinguish the model from usual LRSM, such undesirable couplings are naturally absent so we can have LR symmetry breaking as low as a few TeV. Therefore, from collider perspective ALRM is more interesting than LRSM. An extra global or gauge symmetry group $U(1)_S$ can be introduced to differentiate the bidoublet from its dual. In this alternative formulation right-handed (RH) down-type quarks and right-handed neutrinos are not part of the $SU(2)^\prime_R$ quark and lepton doublets respectively, but rather $d_R, \nu_R$ are $SU(2)^\prime_R$ singlets. New exotic down-type quarks $d_R^\prime$ and neutrinos $n_R$, also known as scotinos, originating from $E_6$, are part of the usual ALRM RH doublets. Left-handed counterparts of those exotic fermions, $d_L^\prime, n_L$, are the $SU(2)_L$ singlets to preserve the LR symmetry of the theory. 

Turning to discrete symmetries, in the literature a generalised lepton number has been defined in two ways: in the Dark LR model (DLRM), $L=S-T_{3R}$ \cite{Ma:2010us} and in Dark LR model 2 (DLRM2), $L=S+T_{3R}$ \cite{Frank:2019nid}. We have adopted the charge assignments as in DLRM2 here. This in turn allows 
one to introduce a generalized $R$-parity, similar to the one existing in supersymmetry, defined here in a similar way as $(-1)^{3B+L+2s}$ \cite{Khalil:2009nb}. Under this $R$-parity,  all SM quarks, leptons and SM gauge bosons are even. The odd $R$-parity particles are as follows: in the scalar sector, $\chi_R^\pm, \, \phi_1^\pm$ and $\phi_1^0$, in the fermion sector, the scotinos $n_L, n_R$, and the exotic quarks $d^\prime_L, d^\prime_R$, and in the gauge sector, $W_R$. All the rest of the particles in the spectrum are $R$-parity even. The existence of a dark matter sector arising from $R$-parity odd particles is another attractive feature of this model, and an advantage over the more common LRSM \cite{Frank:2022tbm}. As such, in this scenario dark matter can be either fermionic (scotinos) or bosonic (CP even or CP odd neutral Higgs bosons).

The scalar sector of the theory consists of one bidoublet $\Phi$ and two Higgs doublets $\chi_{L,R}$. The duals of these scalar fields can be written as $\tilde{\Phi} \equiv \sigma_2 \Phi^\ast \sigma_2$ and $\tilde{\chi}_{L,R} = i\sigma_2 \chi_{L,R}^\ast$ with $\sigma^2$ the Pauli matrix. The symmetry breaking chain from this gauge group to the SM occurs at some LR breaking scale $M_R$, and is effected by vacuum expectation value (\textit{vev}) of RH Higgs doublet $\chi_R$. Subsequent breaking of the SM to low-energy electromagnetic theory is triggered by $\Phi$ and $\chi_L$ \textit{vev}s. In presence of extra $S$ symmetry, quark doublets can interact with $\tilde{\Phi}$ and lepton doublets with $\Phi$ only. With this differentiation one avoids the unwanted mixing between $W_L-W_R$ gauge bosons, as well as between $d, d^\prime$ and $n, \nu$ respectively, in agreement with $R$-parity conservation. The particle content of the model is tabulated in Table \ref{tab:ALRM}. The chosen fermions can constitute $\textbf{27}$ representation of $E_6$.

\begin{table}[htb]
\centering
\begin{tabular}{|c|c|c|c|c|c|c|}
\hline
  & Particles & $SU(3)_C$ & $SU(2)_L$ & $SU(2)^\prime_{R}$ & $U(1)_{B-L}$ & S
           \\[2mm]
\hline \hline 
Quarks & $Q_{L} = 
\begin{pmatrix}
u_L \\
d_L
\end{pmatrix}$ & 3 & 2 & 1 & $\frac{1}{6}$ & 0
          \\[2mm]
         & $Q_{R} = 
\begin{pmatrix}
u_R \\
d'_R
\end{pmatrix}$ & 3 & 1 & $2$ & $\frac{1}{6}$ & $-\frac{1}{2}$
          \\[2mm]
& $d'_L$ & 3 & 1 & 1 & $-\frac{1}{3}$ & $-1$
          \\[2mm]
& $d_R$ & 3 & 1 & 1 & $-\frac{1}{3}$ & 0
          \\[2mm]
\hline
Leptons & $L_{L} = 
\begin{pmatrix}
\nu_L \\
e_L
\end{pmatrix}$ & 1 & 2 & 1 & $-\frac{1}{2}$ & 1
          \\[2mm]
         & $L_{R} = 
\begin{pmatrix}
n_R \\
e_R
\end{pmatrix}$  & 1 & 1 & 2 & $-\frac{1}{2}$ & $\frac{3}{2}$
          \\[2mm]
& $n_L$ & 1 & 1 & 1 & 0 & 2
          \\[2mm]
& $\nu_R$ & 1 & 1 & 1 & 0 & 1
          \\[2mm]
\hline
Scalars & $\Phi = 
\begin{pmatrix}
\phi_1^0 & \phi_1^+\\
 \phi_2^- & \phi_2^0
\end{pmatrix}$ & 1 & 2 & $ 2^{\ast} $ & 0 & $-\frac{1}{2}$
          \\[2mm]
        & $\chi_{L} = 
\begin{pmatrix}
\chi_L^+ \\
\chi_L^0
\end{pmatrix}$ & 1 & 2 & 1 & $\frac{1}{2}$ & 0
        \\[2mm]
        & $\chi_{R} = 
\begin{pmatrix}
\chi_R^+ \\
\chi_R^0
\end{pmatrix}$ & 1 & 1 & 2 & $\frac{1}{2}$ & $\frac{1}{2}$
        \\[2mm]
  \hline
\end{tabular}
\caption{The particle content of ALRM.}
\label{tab:ALRM}
\end{table}

The relevant Yukawa Lagrangian can be written as :
\begin{align}
    \mathcal{L}_Y & = \bar Q_L {\bf Y}^q \tilde\Phi Q_R
    + \bar Q_L {\bf Y}^q_L \chi_L d_R
    + \bar Q_R {\bf Y}^{q}_R\chi_R d'_L
    + \bar L_L {\bf Y}^\ell \Phi L_R
    + \bar L_L {\bf Y}^\ell_L \tilde\chi_L \nu_R
    + \bar L_R {\bf Y}^\ell_R \tilde\chi_R n_L + { h.c.}
\end{align}
Here the Yukawa couplings ${\bf Y}$ are $3 \times 3$ matrices with generation labels as indices. Now, the $\textit{vev}$s of the scalar fields are given by
\begin{equation}
    \langle \Phi \rangle = \frac{1}{\sqrt{2}} \begin{pmatrix}
    0 & 0 \\
    0 & k
    \end{pmatrix}, 
    \langle \chi_L \rangle = \frac{1}{\sqrt{2}} \begin{pmatrix}
    0  \\
    v_L
    \end{pmatrix},
    \langle \chi_R \rangle = \frac{1}{\sqrt{2}} \begin{pmatrix}
    0 \\
    v_R
    \end{pmatrix}.
\end{equation}
The  breaking of the left-right symmetry generates masses for the gauge bosons and, from the Higgs-boson kinetic terms, is responsible for their mixing. The charged gauge bosons do not mix as $\langle\phi_1^0
\rangle = 0$, and their masses are given by
\begin{equation}
  M_{W_L} = \frac12 g_L \sqrt{k^2+v_L^2} \equiv \frac12 g_L v
 \qquad\text{and}\qquad
  M_{W_R} = \frac12 g_R \sqrt{k^2+v_R^2} \equiv \frac12 g_R v' \ .
\label{eq:mw_mwp}
\end{equation}
with $v \equiv \sqrt{k^2 + v_L^2}$ and $v^\prime \equiv \sqrt{k^2 + v_R^2}$. In the neutral sector, the mass squared matrix of the gauge bosons can be written, in the
$\{B_\mu, W_{L\mu}^3, W_{R\mu}^3\}$ basis, as
\begin{equation}
  ({\cal M}^0_V)^2 = \frac14 \bpm
    g_{B-L}^2\ (v_L^2+v_R^2)  & -g_{B-L}\ g_L\ v_L^2    & -g_{B-L}\ g_R\ v_R^2\\
   -g_{B-L}\ g_L\ v_L^2       &  g_L^2\ v^2          & -g_L\ g_R\ k^2\\
   -g_{B-L}\ g_R\ v_R^2       & -g_L\ g_R\ k^2       &  g_R^2\ v^{\prime 2}
  \epm \ ,
\end{equation}
where $g_L$, $g_R$ and $g_{B-L}$ are the gauge coupling constants of  $SU(2)_L$, $SU(2)^\prime_R$ and $U(1)_{B-L}$, respectively. This matrix can be diagonalised through three rotations that mix the $B$, $W_L^3$ and $W_R^3$ bosons into the massless photon $A$ and massive $Z$ and $Z'$ states,
\renewcommand{\arraystretch}{1.}
\begin{equation}
  \bpm B_\mu\\ W_{L\mu}^3\\ W_{R\mu}^3\epm = 
  \bpm \cos \phi_W & 0 & -\sin \phi_W \\ 0 & 1 & 0\\ \sin \phi_W  & 0 & \cos \phi_W  \epm
  \bpm \cos{\theta_W} & -\sin{\theta_W}& 0\\ \sin{\theta_W} & \cos{\theta_W} & 0\\ 0 & 0 & 1 \epm
  \bpm 1 & 0 & 0\\ 0 & \cos{\zeta_W} & -\sin{\zeta_W}\\ 0 & \sin{\zeta_W} & \cos{\zeta_W} \epm
  \bpm A_\mu\\ Z_\mu\\ Z^\prime_\mu\epm  \ ,
\end{equation}
The $\phi_W$-rotation mixes the $B$ and $W_R^3$ bosons into the hypercharge boson $B'$, generated by the breaking of $SU(2)^\prime_{R}\otimes U_{B-L}$ into $U(1)_Y$. The $\theta_W$-rotation denotes the usual electroweak mixing, and the $\zeta_W$-rotation is related to the strongly constrained $Z-Z'$ mixing. The various mixing angles are defined by:
\begin{eqnarray}
  \sin \phi_W &=& \frac{g_{B-L}}{\sqrt{g_{B-L}^2+g_R^2}} = \frac{g_Y}{g_R}
   \qquad\text{,}\qquad
   \sin\theta_W  = \frac{g_Y}{\sqrt{g_L^2+g_Y^2}} = \frac{e}{g_L}     \qquad\text{and}  \nonumber \\
    \tan(2\zeta_W) &=&\frac{2 \cos{\phi_W} \cos{\theta_W} g_L g_R (\cos{ \phi_W}^2 k^2-\sin{ \phi_W}^2 v_L^2) }
     {-(g_L^2 - \cos{\phi_W}^2 \cos{\theta_W}^2 g_R^2) \cos{ \phi_W}^2 k^2 -
         (g_L^2 - \cos{\theta_W}^2 g_{B-L}^2 \sin{ \phi_W}^2) \cos{ \phi_W}^2 v_L^2 +
          \cos{\theta_W}^2 g_R^2 v_R^2} \ .
\label{eq:ewmix}
\end{eqnarray}
Neglecting the $Z-Z'$ mixing, the $Z$ and $Z'$ boson masses are given by
\begin{equation}
  M_{Z} =  \frac{g_L}{2 \cos\theta_W} \ v
  \qquad\text{and}\qquad
  M_{Z'} = \frac{1}{2} \sqrt{ g_{B-L}^2 \sin\phi_W^2 v_L^2 +
     \frac{ g_R^2 (\cos\phi_W^4 k^2 + v_R^2) } {\cos \phi_W^2} 
     } \ .
\label{eq:mz_mzp}
\end{equation}
We note that while the $Z^\prime$ boson interacts with quarks and leptons, $W_R$ couples to quark-exotic quark and charged lepton-scotino. Thus the latter mass is unrestricted by experiment, while the former is expected to be $\ge 4.5$ TeV, from restrictions on its production followed by dilepton decays at the LHC \cite{CMS:2018ipm,ATLAS:2017fih}. The charged $W_R$ boson mass is indirectly restricted from the mass ratio $M_{W_R}/M_{Z^\prime}$, but this constraint is weaker as compared to that for the LRSM. 

The eight degrees of freedom in the charged  scalar sector of the unbroken symmetry, $\phi_1^\pm$, $\phi_2^\pm$, $\chi_L^\pm$ and $\chi_R^\pm$ mix into two physical massive charged Higgs bosons $H_1^\pm$ and $H_2^\pm$, and two massless Goldstone bosons $G_1^\pm$ and $G_2^\pm$ that are absorbed by the $W_L^\pm$ and $W_R^\pm$ gauge bosons,
\begin{eqnarray}
 \bpm \phi_1^\pm\\\chi_L^\pm\epm =
 \bpm \cos\beta & \sin\beta\\ -\sin\beta & \cos\beta\epm
 \bpm H_1^\pm\\G_1^\pm\epm \ , \ \
 \bpm \phi_2^\pm\\\chi_R^\pm\epm =
 \bpm \cos\zeta & \sin\zeta\\ -\sin\zeta & \cos\zeta\epm
 \bpm H_2^\pm\\G_2^\pm\epm \ ,
\label{eq:ch_mix}
\end{eqnarray}
with
\begin{equation}
  \tan\beta = \frac{k}{v_L} \qquad\text{and}\qquad \tan\zeta = \frac{k}{v_R} \ .
\end{equation}
The masses of the new physical charged bosons, relevant for our study, are given in terms of the parameters of the Lagrangian and the \textit{vev}s, as
\begin{equation}
M_{H_1^\pm}=v^2~(\alpha_3-\alpha_2)-\frac{\mu_3}{\sqrt{2}}~\frac{v^2v_R}{kv_L},~~~~~~{\rm and},~~~~~~~
M_{H_2^\pm}=(k^2+v_R^2)~(\alpha_3-\alpha_2)-\frac{\mu_3}{\sqrt{2}}~\frac{(k^2+v_R^2)v_L}{kv_R} \, .
\end{equation}
 where $H_1^\pm$ is $R$-parity even, while  $H_2^\pm$ is $R$-parity odd.  The masses of the exotic fermions, for each generation, are given by
\begin{equation}
    M_n = \frac{1}{\sqrt{2}}{\bf Y}^\ell_R~v_R,~~~~M_{d'} = \frac{1}{\sqrt{2}}{\bf Y}^q_R~v_R
\end{equation}
These Yukawa couplings are not connected with masses of other fermions, and therefore, can be varied independently, with the maximum limited by perturbativity. Active neutrinos get their mass through see mechanism with the help of $\nu_R$. 
The flavour eigenstates $\nu_{\alpha L}$, $\nu_{\alpha R}$ and $n_\alpha$ can be expressed in terms of admixture of mass eigenstates ($\nu_i$, $N_j$, $n_k$) as follows,
\begin{eqnarray}
\nu_{\alpha L} &=& U_{\alpha i}^{\nu \nu} \nu_i + V_{\alpha j}^{\nu N} N_j
\label{eq:LHmassbasis}\\
\nu_{\alpha R} &= &S_{\alpha i}^{N \nu} \nu_i + T_{\alpha j}^{N N} N_j
\label{eq:RHmassbasis}\\
n_\alpha &=& A_{\alpha k}n_k
\end{eqnarray}
where $i,j,k =1,2,3$, and $U,~V,~S,~T$ and $A$ are the corresponding mixing matrices. For simplicity, we shall assume diagonal ${\bf Y}^\ell_R$ so that $A$ is identity matrix. We shall also assume that the active-sterile mixing is negligible. 

In what follows we explore production and decay processes at the LHC, with the aim of observing an imprint of $W_R$ bosons at the colliders.

\section{Collider Analysis}
\label{sec:collider}

The dominant production and decay channels of the $W_R$ boson depend upon the mass spectrum of the model.  $W_R$ couples to the right-handed fermions, $u_R,~d_R',~n_R,~e_R$, the charged Higgs, $H_1^\pm,~H_2^\pm$, the SM Higgs, $h$,  the two heavier neutral scalars,  the pseudoscalar, and the neutral gauge bosons $\gamma, ~Z, ~Z'$. With the expressions for masses given in the previous section, we can distinguish different mass hierarchies.  In this study, $H_1^\pm$ is taken to be much heavier than other particle, especially $W_R$ and $H_2^\pm$, and therefore do not involve in our analysis.

Based on the mass ordering of the three exotic down-type quarks, $d'$ and scotinos, $n$ (with the $d'$ and $n$ representing all the three flavours),  we divide our study into different cases. In the first instance (case {\bf 1}) we assume that the exotic quarks are lighter than $W_R$, but heavier than $H_2^\pm$. The exotic scotino is taken to be lighter than the charged Higgs. That corresponds to the mass hierarchy
\begin{equation}
    m_{n_i}< M_{H_2^\pm} < m_{d'}< M_{W_R}
\end{equation}
The lightest of the scotinos is the lightest dark sector particles in this case, and therefore the dark matter candidate. In the second scenario, we analyze  the case where the exotic quarks are heavier than the $W_R$ bosons, but the scotinos are lighter than $W_R$. Within this, two different possibilities of charged Higgs masses are considered. In one we assume that the dark scalars, including the charged scalars, are lighter than the scotinos. We call this case {\bf 2a}, with mass hierarchy
\begin{equation}
     M_{H_2^\pm} < m_{n_i}< M_{W_R}< m_{d'}
\end{equation}
Considering the reverse hierarchy between $H_2^\pm$ and $n_i$, we have case {\bf 2b} with
\begin{equation}
      m_{n_i}< M_{H_2^\pm} <M_{W_R}< m_{d'}
\end{equation}
In case {\bf 2a} the lightest dark sector particles are the degenerate $H_1^0$ and $A_1^0$, while in case {\bf 2b} the dark matter candidate is again the lightest scotino. As a third case we consider both the scotinos and exotic quarks heavier than both $W_R$ and $H_2^\pm$ (case {\bf 3}):
\begin{equation}
       M_{H_2^\pm} <M_{W_R}< (m_{d'}, m_{n}),
\end{equation}
where $d'$ stands for all the down type exotic quarks, and $n$ stands for all the scotinos. Each of these cases lead to distinct decay channels for $W_R$ and $H_2^\pm$. Thus, in our  study of production and decay of $W_R$ and $H_2^\pm$, these different cases provide distinct final states.

We now proceed, along the hierarchies described above, with our detailed analysis of the pair production of $W_R$ and $H_2^\pm$ as well as their associated production. As the cross section for these process at LHC  at 13 TeV is in general small, we evaluate {\it all} signals and corresponding backgrounds at a 27 TeV collider, and at 13 TeV only when promising.
Rather than scanning over the large parameter space of the ALRM, we focus on target regions of the parameter space where the signals may be promising. In this case, for a specific example we have chosen a benchmark featuring common set of parameter for processes {\bf 1a} and {\bf 1b} as given in Table \ref{tab:parameters1}. Table \ref{tab:parameters1} also contains masses of the extra gauge bosons, charged Higgs and all the exotic quarks. With the given $g_R,~v'$ and $\tan\beta$, the $W_R$ has a mass of 1146 GeV without violating any experimental constraints. The $pp\to W_RW_R$ production cross section with in this case is, therefore, 0.76 pb. The production cross sections at this centre of mass energy are given in Table~\ref{tab:productioncs}.
\begin{table}[htb!]
\caption{\label{tab:parameters1}Parameter values for processes {\bf 1a} and {\bf 1b}. We list, in the first row, the parameter in the scalar potential, in the next row, the masses for the exotic quarks and leptons, in the following rows the masses obtained for the additional gauge bosons and the relevant charged, $H_2^\pm$ and the production cross sections for processes {\bf 1a} and {\bf 1b} at 27 TeV. The last two rows list all the relevant branching ratios for $W_R$ and $H_2^\pm$ yielding the final states. We have assumed in both cases, the minimum $p_T$ for both jets and leptons to be 10 GeV.}
  \begin{center}
 \small
 \begin{tabular*}{1.0\textwidth}{@{\extracolsep{\fill}}ccccccc}
 \hline\hline
	$\lambda_2$ & $\lambda_3$ &$\alpha_1=\alpha_2=\alpha_3$ &$\mu_3$ &$\tan \beta$ &$g_R$ &$v^\prime$ 
	\\
	\hline
	0.5 & 1.6 & 0.01 &-400 GeV & 50 & 0.37 & 6.2 TeV \\
  \hline\hline   
 $ m_{d^\prime}$ & $m_{s^\prime}$ & $m_{b^\prime}$ & $m_{n_e} $ & $m_{n_\mu}$ & $m_{n_\tau}$
\\ 
	 \hline
 600 GeV & 700 GeV & 800 GeV & 140 GeV & 135 GeV & 130 GeV	 \\ 
\hline \hline
$M_{W_R}$ & $M_{Z^\prime}$ &  $M_{H_2^-}$ &$M_{H_1^0}=M_{A_1^0}$ &  cross section {\bf 1a} & cross section {\bf 1b}  \\	 
\hline
 1.15 TeV & 4.55 TeV &187 GeV&309 GeV & $5.59$ fb & $0.087$ fb \\	 
  \hline \hline
  BR($W_R \rightarrow u d^\prime$) & BR($W_R \rightarrow c s^\prime$) & BR($W_R \rightarrow t b^\prime$) & BR($W_R \rightarrow e n_e$) & BR($W_R \rightarrow \mu n_\mu$)& BR($W_R \rightarrow \tau n_\tau$)\\ 
  \hline
  23.70 \%& 18.43 \%& 11.77 \% & 9.33 \% & 7.65 \% & 5.86 \% \\ 
  \hline \hline
  BR($H_2^- \rightarrow \tau n_\tau$) & BR($H_2^- \rightarrow \tau n_\mu$)& BR($H_2^- \rightarrow \tau n_e$) &BR($H_2^- \rightarrow e n_e$) &&\\ 
  \hline
  59.58 \% & 25.31 \% & 12.23 \% & 0.81 \% &&\\ 
  \hline \hline
    \end{tabular*}
\end{center}
 \end{table}

\begin{table}
\caption{ Production cross section of $W_RW_R$, $W_RH_2^\pm$ and $H_2^+H_2^-$ at $\sqrt{s}=27$ TeV LHC.
}
\vskip 3mm
\label{tab:productioncs}
\begin{tabular}{l|c|c|c}
\hline\hline
&&&\\
process&~$pp\to W_RW_R$~&~$pp\to W_RH_2^\pm$~&~$pp\to H_2^+H_2^-$~\\[2mm]\cline{1-4}
&&&\\
cross section (fb)&760&2.6&250\\[2mm]
\hline\hline
\end{tabular}
\end{table}

\subsection{Case 1:  $\mathbf {M_{W_R} >m_{d^\prime}, m_{s^\prime}, m_{b^\prime}}$}     %
\label{subsec:1}												%

In this subsection we investigate the pair production of $W_R$ bosons ({case {\bf 1a}}), as well as the $W_R H_2^\pm$ associated production ({case {\bf 1b}}), for the case where the exotic quarks are lighter than the $W_R$ boson. 

\subsubsection{Case {\bf (1a)} ~~$ pp \rightarrow W_R W_R,~ W_R \rightarrow u d^\prime,~ d^\prime \rightarrow H_2^- u,~ H_2^- \rightarrow n_\tau \tau$}

\begin{figure}[htb!]
    \centering
\includegraphics[scale=0.45]{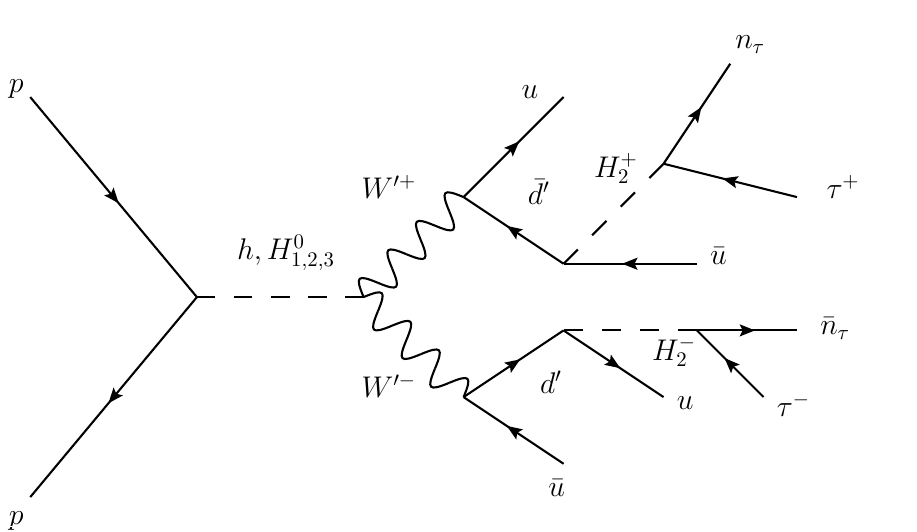}\qquad
\includegraphics[scale=0.45]{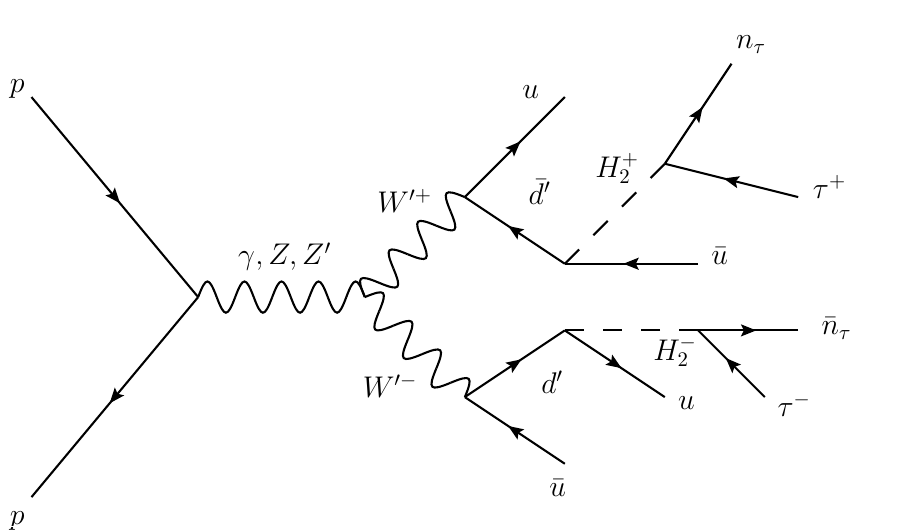}\qquad
\includegraphics[scale=0.3]{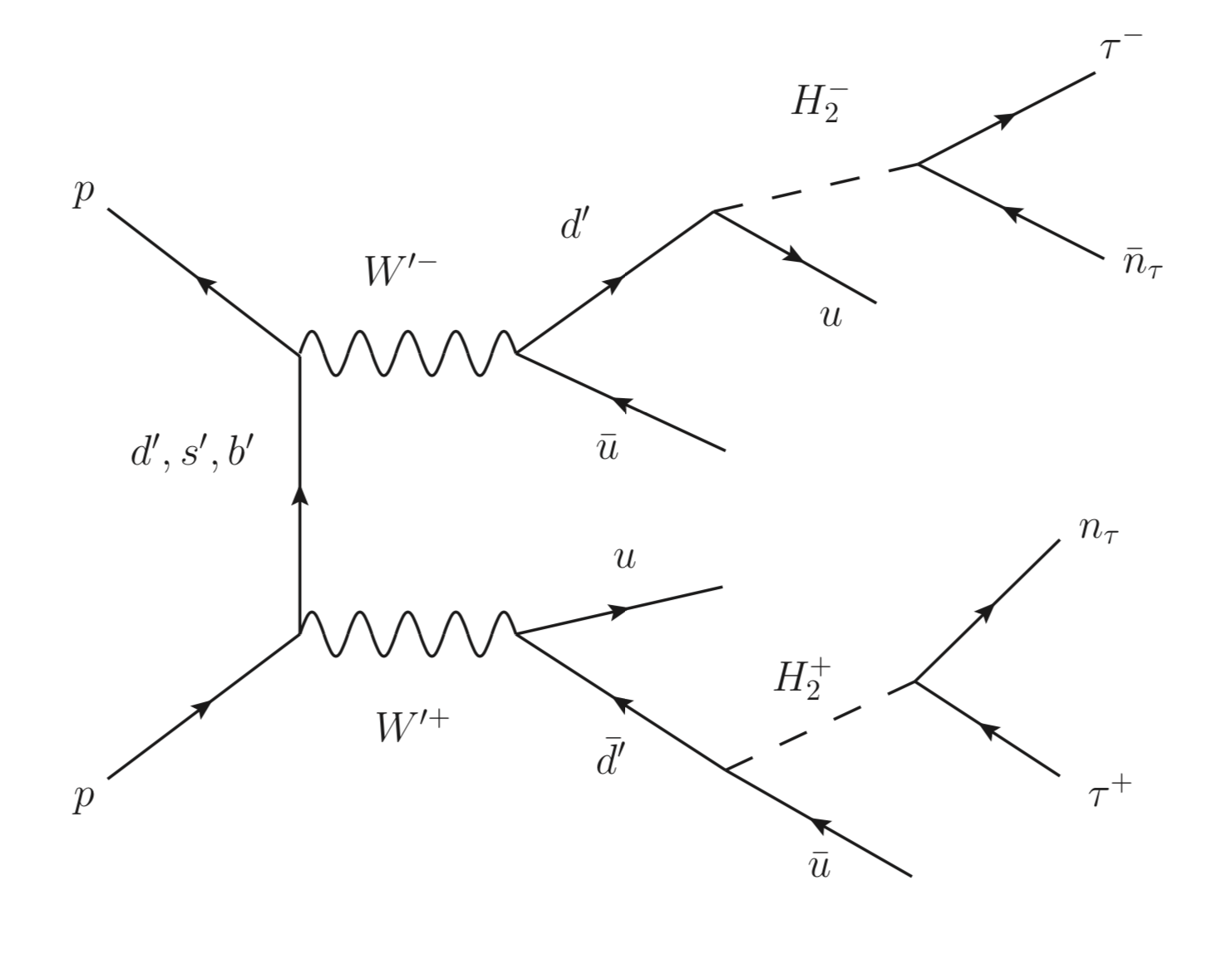}
\caption{Production and decay processes for {\bf 1a}, through both the $s$ (top row) and $t$ (bottom row) channels.}
    \label{fig:feyn1a}
\end{figure}

Here $u$ stands for the light quarks, $u$ and $c$, and $d'$ for $d'$ and $s'$. We plot the Feynman diagram responsible for this process in Fig. \ref{fig:feyn1a}, for both the $s$ and $t$ channels. The final state contains 4 jets, two taus and two scotinos (appearing as missing energy) ($4j+2\tau+$MET). Considering large jet multiplicity, we do not analyse this process any further in our parton-level study.

\subsubsection{Case {\bf (1b)} ~~$
pp \rightarrow W_R H_2^+,~ W_R \rightarrow u d^\prime,~ d^\prime \rightarrow H_2^- u,~ H_2^- \rightarrow n_\tau \tau$}

\begin{figure}[htb!]
    \centering
\includegraphics[scale=0.4]{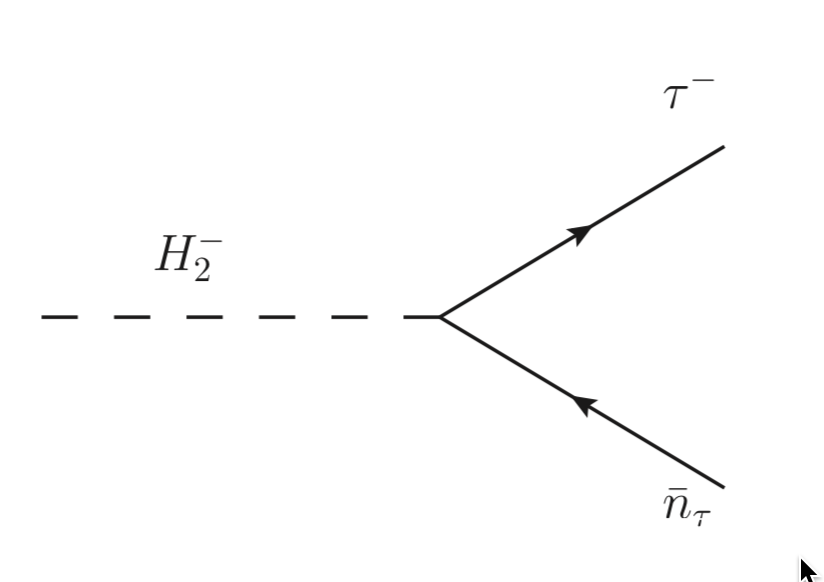}\qquad
\includegraphics[scale=0.3]{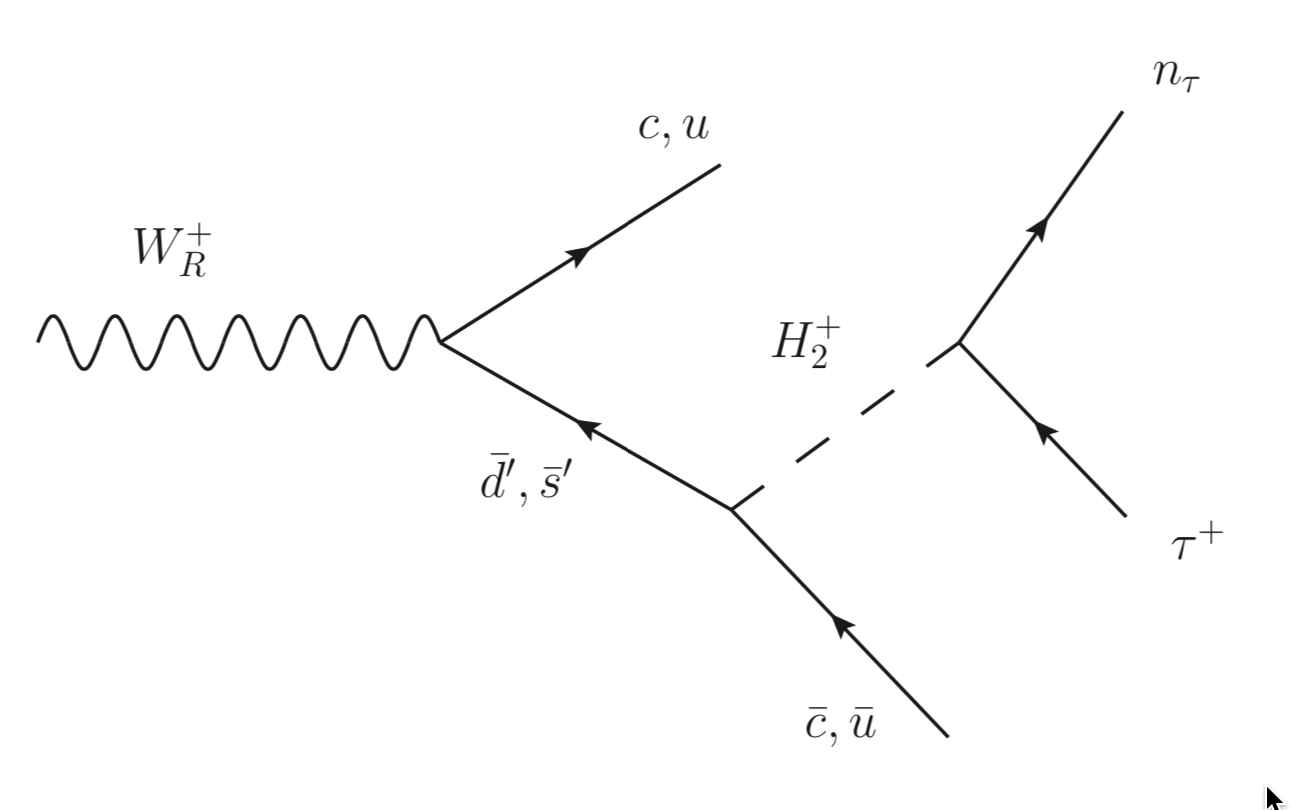}
    \caption{Decay of $H_2^-$ and $W_R$ as described in process {\bf{1b}}.}
    \label{fig:feyn1b}
\end{figure}

In Fig. \ref{fig:feyn1b} we plot the Feynman diagram responsible for $H_2^\pm$ and $W_R$ decay into final state particles for the associated $W_RH_2^\pm$ production corresponding to both the $s$ and $t$ channels where production channels for $H_2^\pm$ and $W_R$ are exactly same as case {\bf (1a)}. The final state contains 2 jets, two taus and two scotinos, appearing as missing energy, ($2j+2\tau+$MET).

Including all the relevant branching fractions as described above,  and summing over all possible channels that lead to $4j+2\tau+$MET, the cross section becomes 5.59 fb. The decay of the $W_R$ boson in this case will be dominantly into the $ud'$ (23.7\%), $cs'$ (18.4\%) and $tb'$ (11.8\%), and with smaller probabilities into leptons and scotinos.  As relevant to the process discussed here, we consider both $W_R\to ud'$ and $W_R\to cs'$ decay channels. With the mass hierarchy set as $m_{d',s',b'} >M_{H_2^\pm}$, the exotic quarks further decay into usual quarks and $H_2^\pm$. Considering the light flavoured quark channels, the decays are $d'\to uH_2^-$ with branching ratio of 21.7\% and $d'\to cH_2^-$ (3.2\%), $s'\to uH_2^-$ (1.7\%) and $s'\to cH_2^-$ (75.3\%). The $R$-parity negative charged Higgs boson $H_2^\pm$ subsequently decay to $n$ (scotino) and a charged lepton, with the tau lepton channel dominating.  

\begin{figure}[htb!]
    \centering
    \includegraphics[scale=0.42]{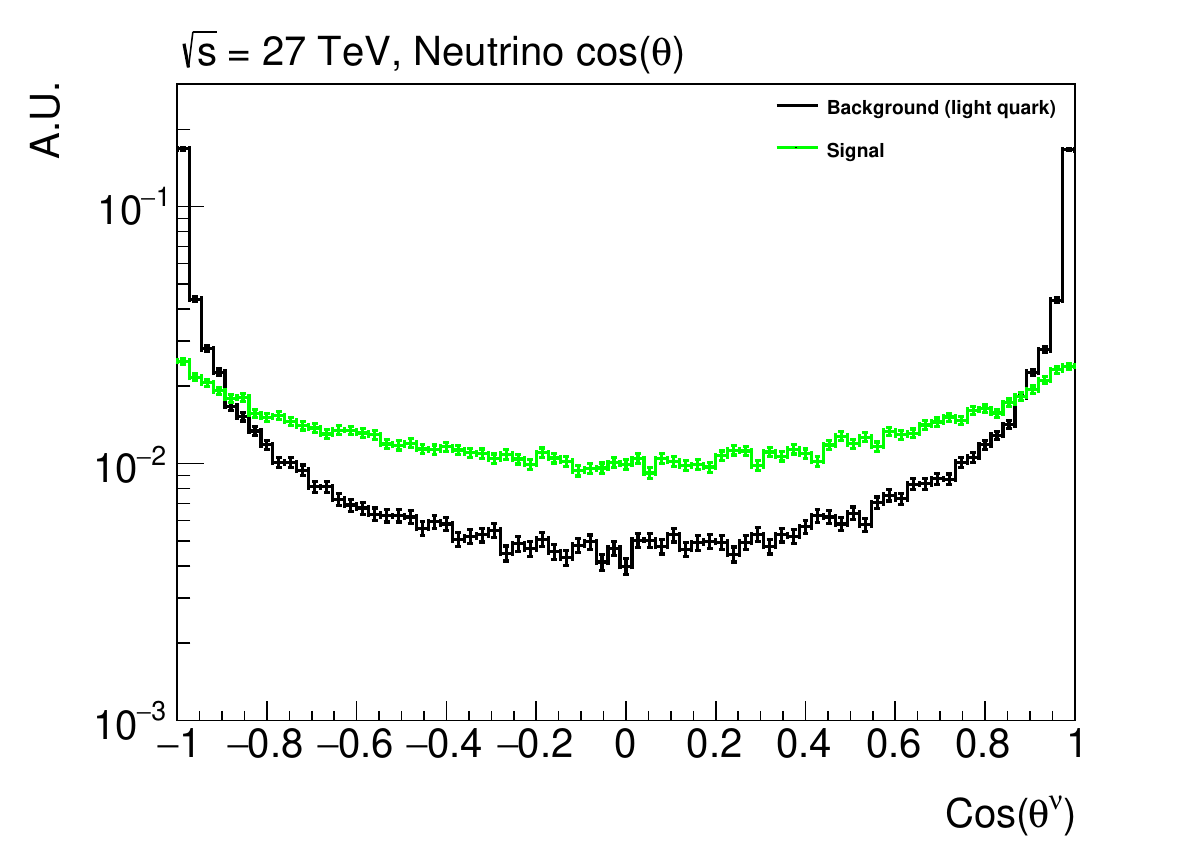} \qquad
    \includegraphics[scale=0.42]{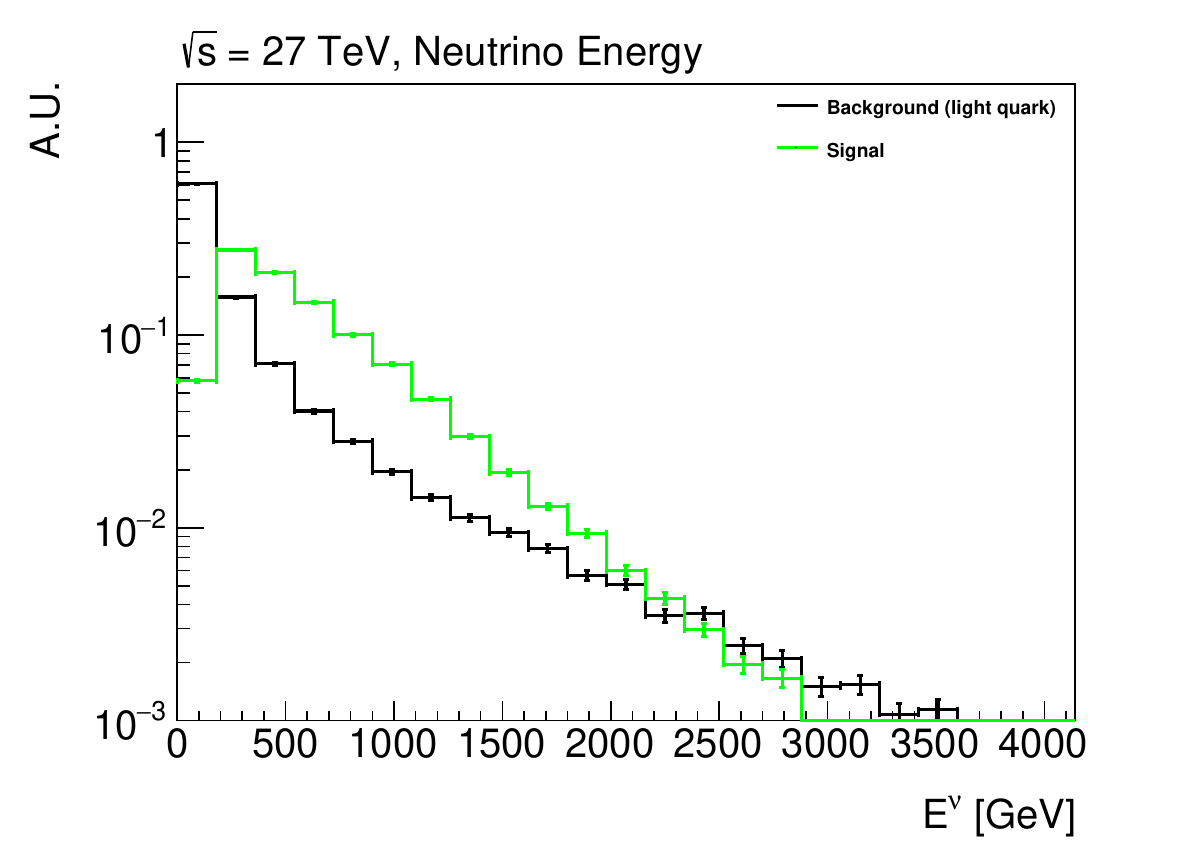}\qquad
    \includegraphics[scale=0.42]{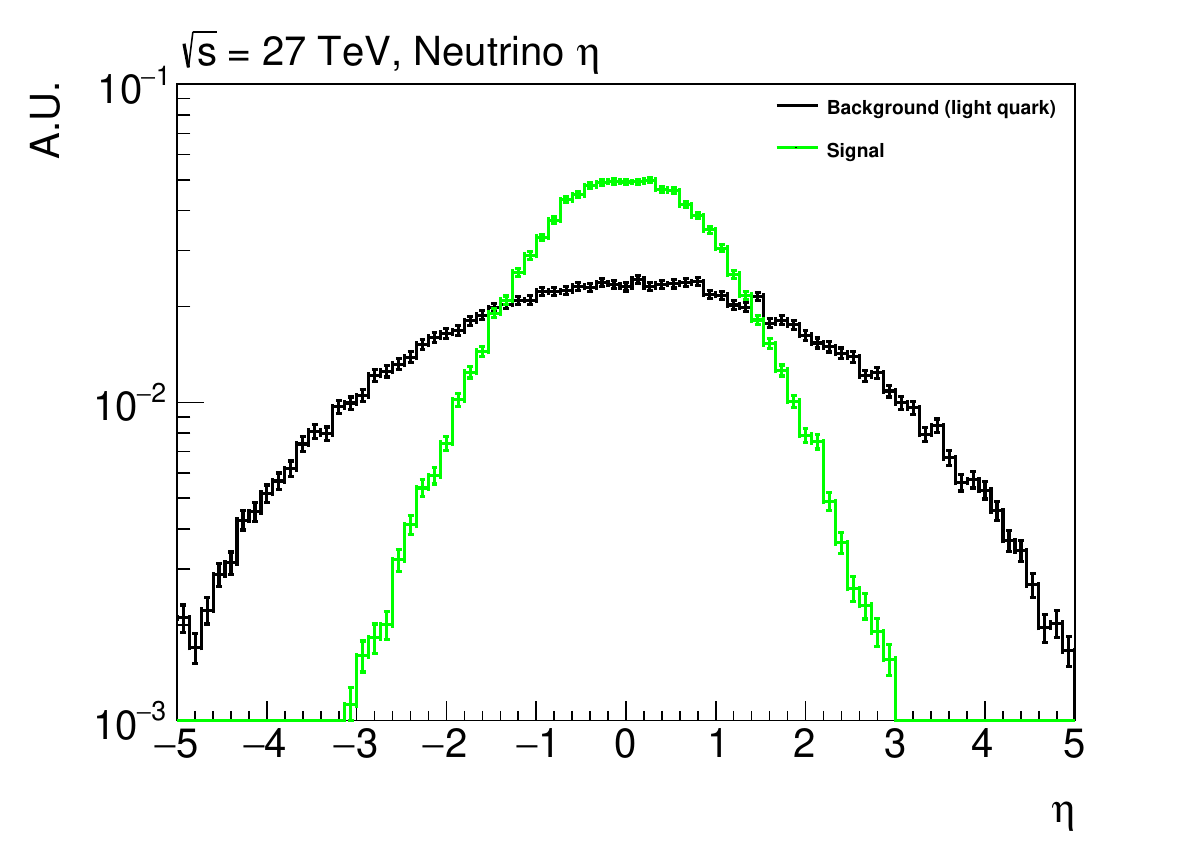}\qquad
    \includegraphics[scale=0.42]{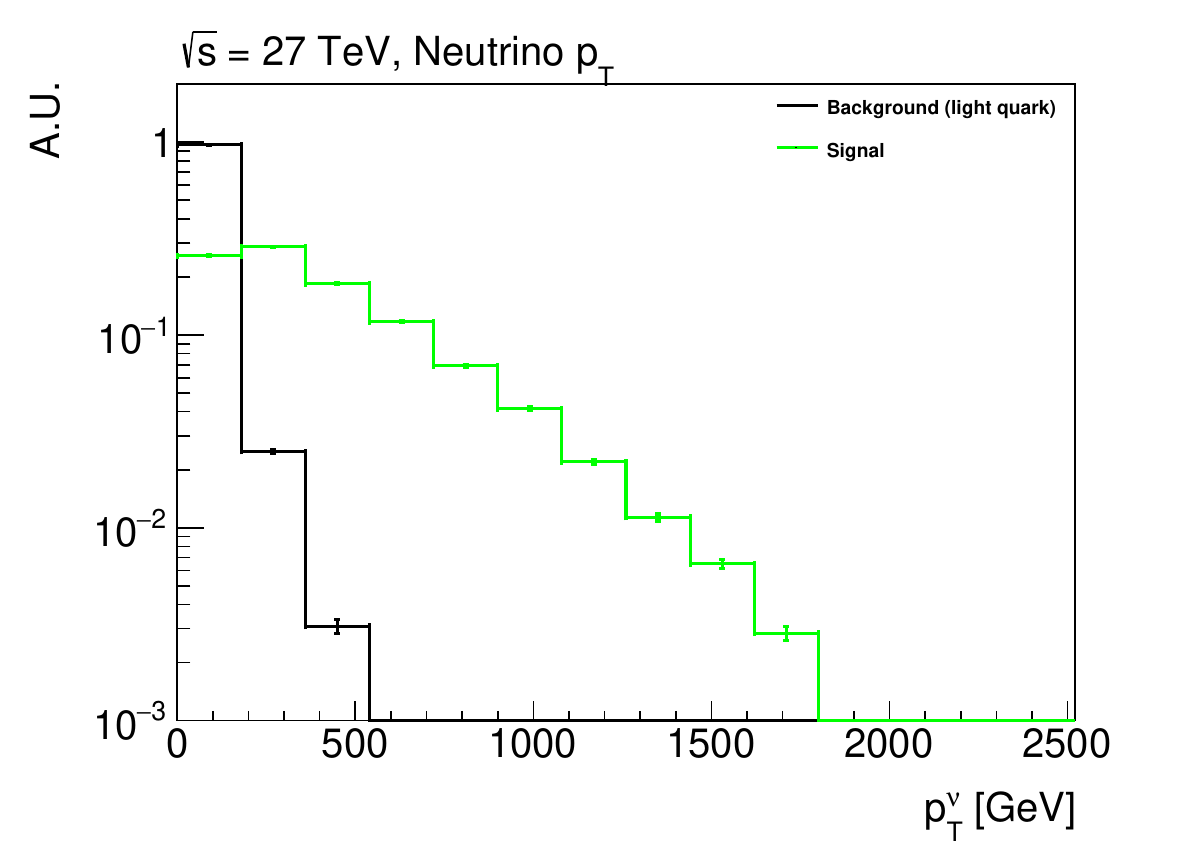}
    \caption{Signal and background for processes  {\bf 1b}. We show the angular separation  between the incoming proton beam and the particle detected as missing energy (scotino or SM neutrino) (top left), the scotino energy (top right), the pseudorapidity (bottom left) and the scotino transverse momentum (bottom right), before cuts. One can see that, in all plots, the signal (in green) rises above the background in some parameter region.}
    \label{fig:hist1b}
\end{figure}

For simulation of the signal events and analysis, we used the {\tt FeynRules-v2.3}  \cite{Alwall:2014bza,Alloul:2013bka,Degrande:2011ua}
implementation of the ALRM as in \cite{Frank:2019nid} and exported the output into {\tt MadGraph5\_aMC} version {\tt 3.2.0} \cite{Alwall:2011uj} to simulate the hard-scattering cross-section for both the signal and background. The signal from {\bf 1a}, containing 4 jets and large missing $p_T$, has practically no significant background from the SM. With a cross section of 5.59 fb, we expect a large number of signal events, even at moderate luminosity, 300 fb$^{-1}$. For process {\bf 1b}, the dominant background comes from $pp \rightarrow j j \tau^+ \tau^- \nu_\tau \bar{\nu}_\tau$ where $j$ corresponds to light quarks i.e.,  $j \equiv u,~ \bar{u},~ d,~\bar{d}, ~c,~\bar{c}, ~s,~\bar{s}$. We did not include the gluon, as the ISR and FSR effects are not considered, and we anticipate a signal selection with large $p_T$. To improve the sensitivity, we select out kinematic regions with larger effect on the signal events. A study of the kinematic distributions as presented in Fig. \ref{fig:hist1b}, and we identify that the best selection criteria are to impose restrictions on the missing energy and missing transverse momentum. 
The cuts and selection for this process are given in Table \ref{tab:cuts1}, where we also present the signal significance obtained. The significance is evaluated using  \cite{bityukov1998observability}
\begin{equation}
\label{eq:sign}
  s    = \frac{S}{\sqrt{B + \sigma_B^2}} \,
\end{equation}
where the number of selected signal and background events are denoted by $S$ and $B$, respectively, and, in addition, we assume an expected error of $\sigma^2_{\rm B}$ in the background.
\begin{table}
    \begin{center}
\caption{\label{tab:cuts1} Imposed cuts on the backgrounds and signal for process  {\bf 1b}, for $\sqrt{s}=27$ TeV \@ 3000 fb$^{-1}$ and the number of events for signal and backgrounds before and after imposing cuts. Here $\theta^{mis}$ is the angular separation  between the tau and missing particle, $E$ is the energy, $p_T$ ~is the transverse momentum, and $\eta$ is the pseudorapidity of the missing particle.  The missing particle is the scotino in the case of signal events, and neutrino in the case of background.}
  \small 
   \begin{tabularx}{\textwidth}{l|c|c|c}
\hline\hline
&&&\\
Selection & Signal events & Background events & Significance \\[1mm]
& ($S$) & ($B$) & ($s$) \\[2mm]
\hline
&&&\\
Before cut: @~~$\sqrt{s}=27$ TeV, ~~$\int{\cal L} = 3000$~fb$^{-1}$ & 261&810000&0.2\\[2mm]
\hline
&&&\\
After selection cuts:&&&\\
 $|\rm{cos}(\theta^{mis})|  < 0.9 $,~ $200$ GeV $<E< 2300 $ GeV,~
 $p_T> 200$ GeV,~~ $|\eta|<1.4$ & 135&10789&1.2\\
\hline\hline
\end{tabularx}
\end{center}
\end{table}
The crude selection criteria we have adopted here still leaves large background events, leading to small signal significance. A more fine tuned selection criteria, along with detector efficiencies are required to find more realistic significance. Overall, seeing the hint of presence of $W_R$ is  not very promising through this channel.

\subsection{Case 2:  $\mathbf {M_{W_R}<m_{d^\prime}, m_{s^\prime}, m_{b^\prime}}$ -- leptonic decays} %
\label{subsec:2}														     %

In this section, we investigate collider events for the case where the exotic quarks are heavier than the $W_R$ boson, thus restricting its decay mostly to the leptonic channels involving the scotinos. We consider two different mass hierarchy for the scotinos. One mass hierarchy is $m_{n_e}<m_{n_\mu}<m_{n_\tau}$ (case {\bf 2a}) and the other has $m_{n_\tau}<m_{n_\mu}<m_{n_e}$ (case {\bf 2b}), as given in Table~\ref{tab:parameters2}. The exotic quark masses are taken be $m_{d'} = 1.3$ TeV,~$m_{s'}=1.4$ TeV, $m_{b'}=1.5$ TeV, heavier than the $W_R$ boson with $M_{W_R}=1146$ GeV, as before (Table \ref{tab:parameters1}). With this, the $W_R$ decays mostly to scotinos and leptons. We shall discuss the cases {\bf 2a} and {\bf 2b} separately below.

\subsubsection{Case {\bf 2a}: $pp\to (W_RW_R+W_RH_2^\pm+H_2^-H_2^+)\to e^+e^- +$ MET }

In case {\bf 2a}, the parameters are chosen so that the $R$-parity odd neutral scalars have masses $M_{H_1^0/A_1^0} = 151$ GeV, while the $R$-parity odd charged scalar maintains mass $M_{H_2^-} = 187$ GeV. The parameter that controls the mass of $H_1^0/A_1^0$ is $\lambda_2$, which does not affect the charged scalar mass. We have considered $\lambda_2=-0.1$ here. With this mass hierarchy, $H_2^+$ decay mostly into the neutral scalars and virtual $W_L$'s, with about 7.7\% branching ratio to $H_2^-\to H_1^0/A_1^0~e \nu$. In this scenario, the neutral scalars $H_1^0$ and $A_1^0$ are the (degenerate) dark matter candidates. At the same time, $W_R$ decays mostly into scotinos and leptons, with $W_R\to n_e e$ having a branching ratio of $26.8\%$. The scotino, $n_e$ further decays into $\nu H_1^0/A_1^0$ with a branching ratio of 25.3\%, after summing over the neutrino flavours. We focus here on $ pp \to W_RW_R,~W_RH_2^\pm$ and $H_2^+H_2^-$ production with the above decays of $W_R$ and $H_2^\pm$ leading to decaying to  $e^+e^-+MET$ in the final state. With the production cross sections given in Table \ref{tab:productioncs}, this leads to a total cross section times branching ratios of 4.96 fb for $pp \to e^+e^-$+MET through the above channels.
 
\begin{figure}[htb!]
   \centering
\includegraphics[scale=0.4]{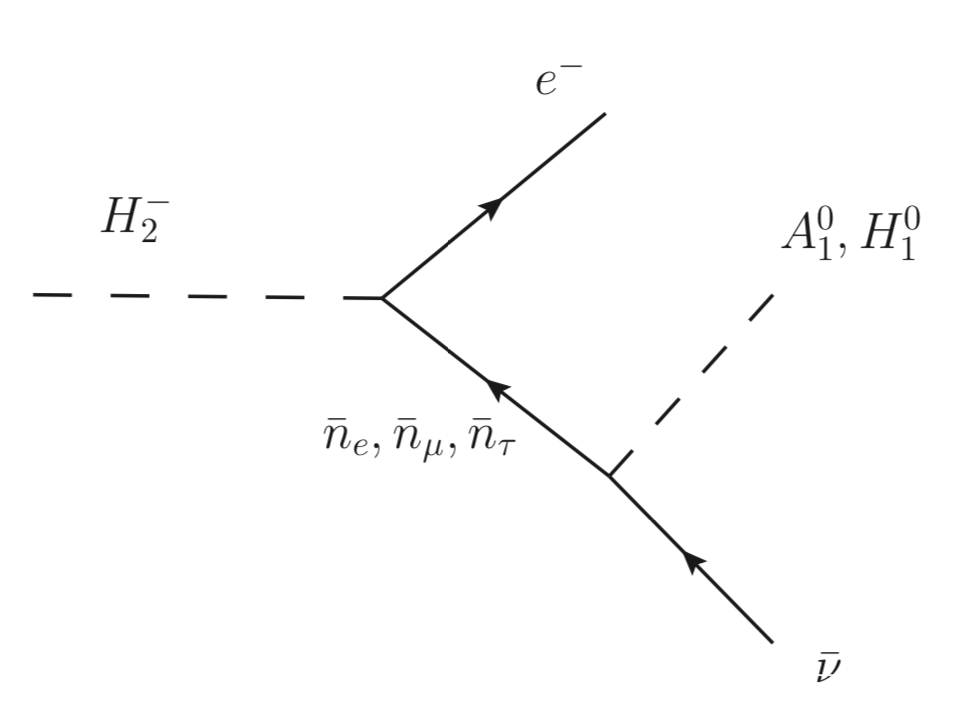}\qquad
\includegraphics[scale=0.4]{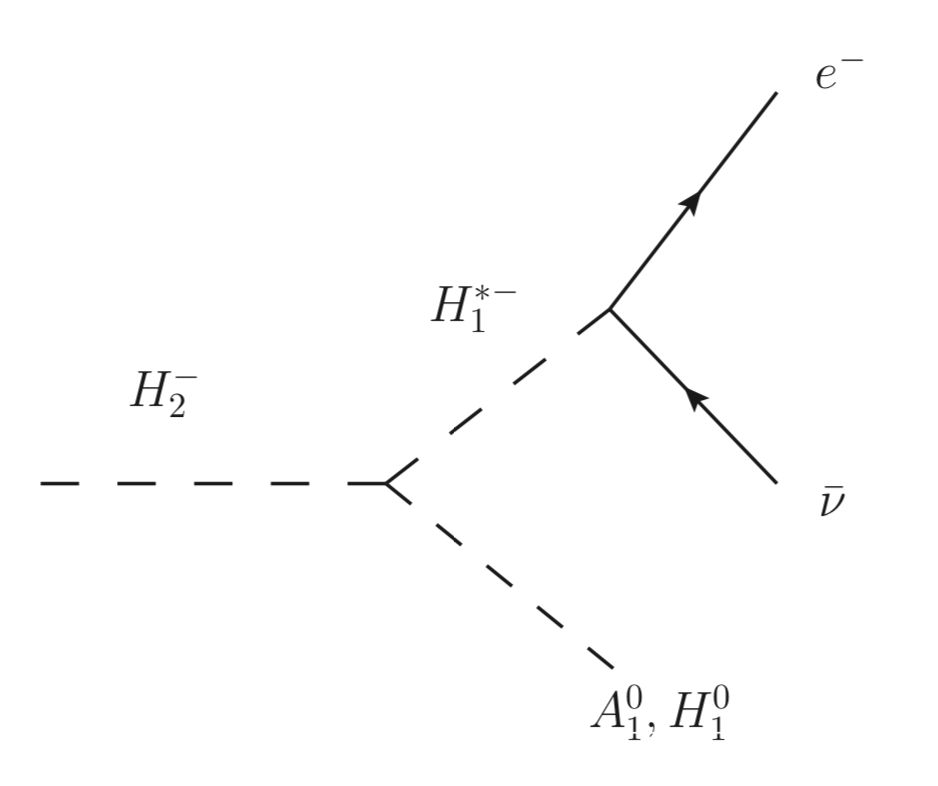}\qquad
\includegraphics[scale=0.35]{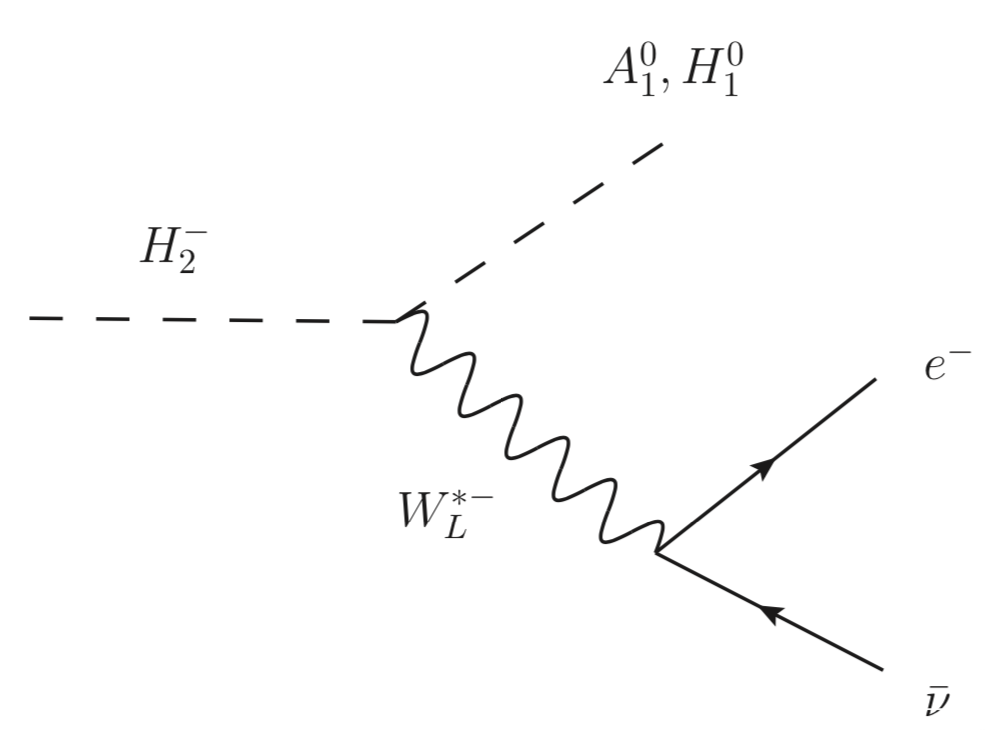}\qquad
\includegraphics[scale=0.35]{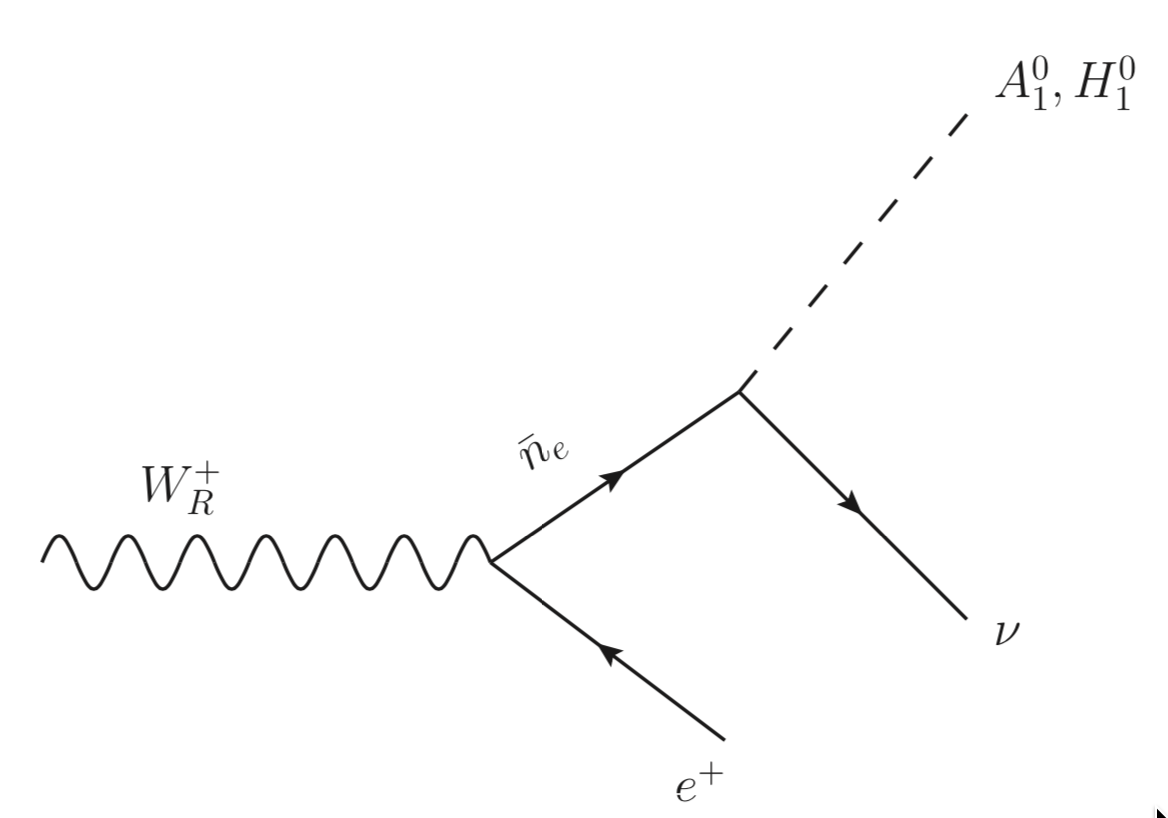}
   \caption{Decay processes of $H_2^-$ and $W_R$ relevant to the process $pp\to (W_RW_R+W_RH_2^\pm+H_2^-H_2^+)\to e^+e^- + $MET for the case {\bf 2a}, with the further invisible decay $n_e, n_\mu, n_\tau \to \nu H_1^0/A_1^0$.}
    \label{fig:feyn2a}
\end{figure}

\begin{table}[htbp]
\caption{\label{tab:parameters2}Parameter values for process {\bf 2a} and {\bf 2b}. We list, in the first row the masses for the scotinos, in the following row the relevant branching ratios for $W_R$ and $H_2^\pm$ yielding the final states, and in the last row the production cross sections times branching ratios for processes {\bf 2a} and {\bf 2b} for $\sqrt{s}=27$ TeV. The rest of the parameters (except for the exotic quarks masses, which play no role here) are as in Table \ref{tab:parameters1}. We have assumed in both cases the minimum $p_T$ for both jets and leptons to be 10 GeV.}
\begin{center}
 \small
 \begin{tabular*}{1.0\textwidth}{@{\extracolsep{\fill}}c|ccc|c|ccc}
 \hline\hline   
 {\bf 2a}& $m_{n_e} $ & $m_{n_\mu}$  & $m_{n_\tau}$ & {\bf 2b} &$m_{n_e} $ & $m_{n_\mu}$ & $m_{n_\tau}$ 
\\ 
	 \hline
 & 300 GeV & 500 GeV & 700 GeV &  & 140 GeV &135 GeV &130 GeV \\
  \hline \hline
  & BR($W_R \rightarrow e n_e$)  & BR($H_2^- \rightarrow H_1^0/A_1^0~e \nu_e$) &BR($n_e \rightarrow H_1^0/A_1^0~\nu$)   
  & & BR($W_R \rightarrow \tau n_\tau$)  & BR($H_2^- \rightarrow \tau n_\tau$)   
  & 
  \\ 
  \hline
 &26.58 \%& 7.73\%
 &25.35 \% 
 &   & 17.47 \% & 59.58\%    
 & 
  \\ 
  \hline \hline  
  && cross section~$\times$ BR && &&~cross section $\times$ BR  \\	
  \hline
 &&    4.96 fb &&&& 108.7 fb\\ 
  \hline \hline
    \end{tabular*}
\end{center}
 \end{table}
The decay of produced $H_2^\pm$ and $W_R$ proceed according to the Feynman diagrams in Fig. \ref{fig:feyn2a}.

The background process is taken as $pp \to e^+e^- + \nu\bar\nu$ where $\nu$ corresponds to all SM neutrino flavors. It is possible that the Drell-Yan process $pp\to e^+e^-$ contributes also to the background once the detector effects are considered. However, we are limiting our study to the parton level, and therefore only backgrounds that mimic the final state at the parton level are considered here.
\begin{figure}[htb!]
    \centering
    \includegraphics[scale=0.4]{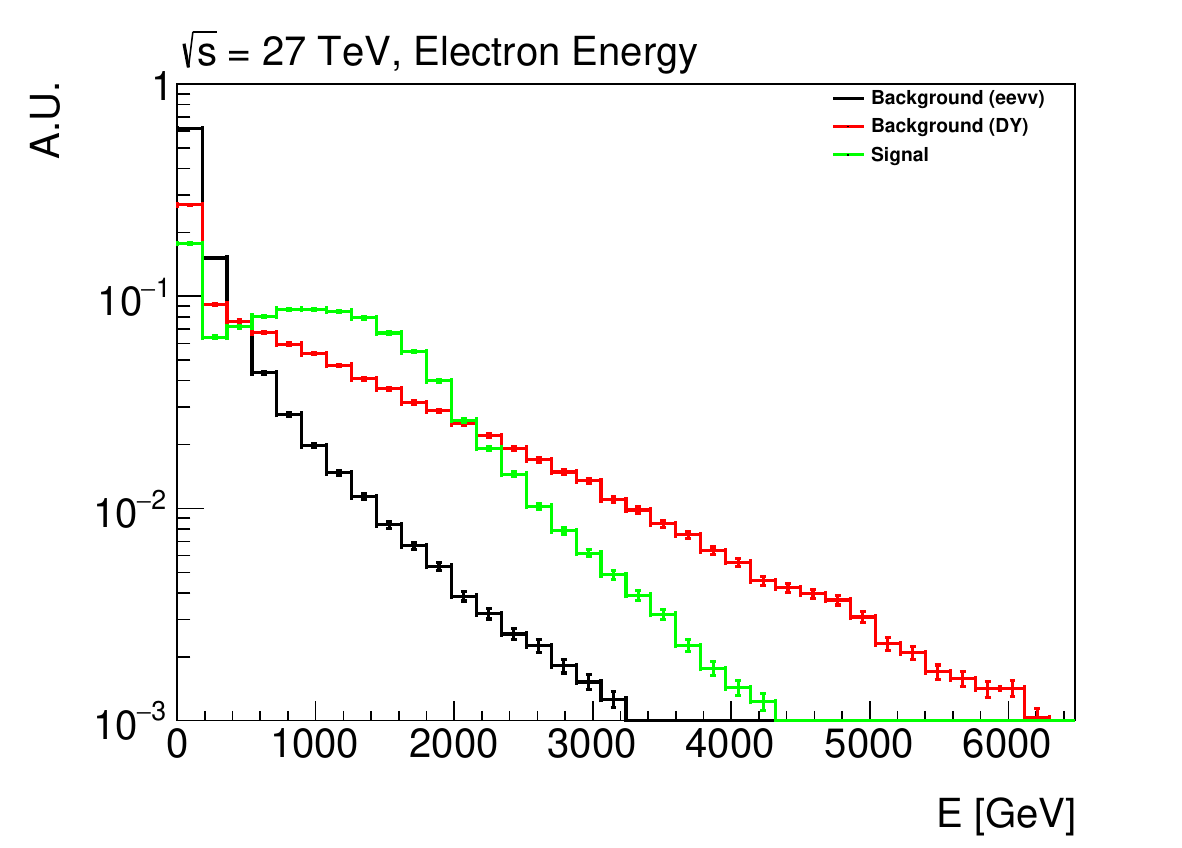}\qquad
    \includegraphics[scale=0.4]{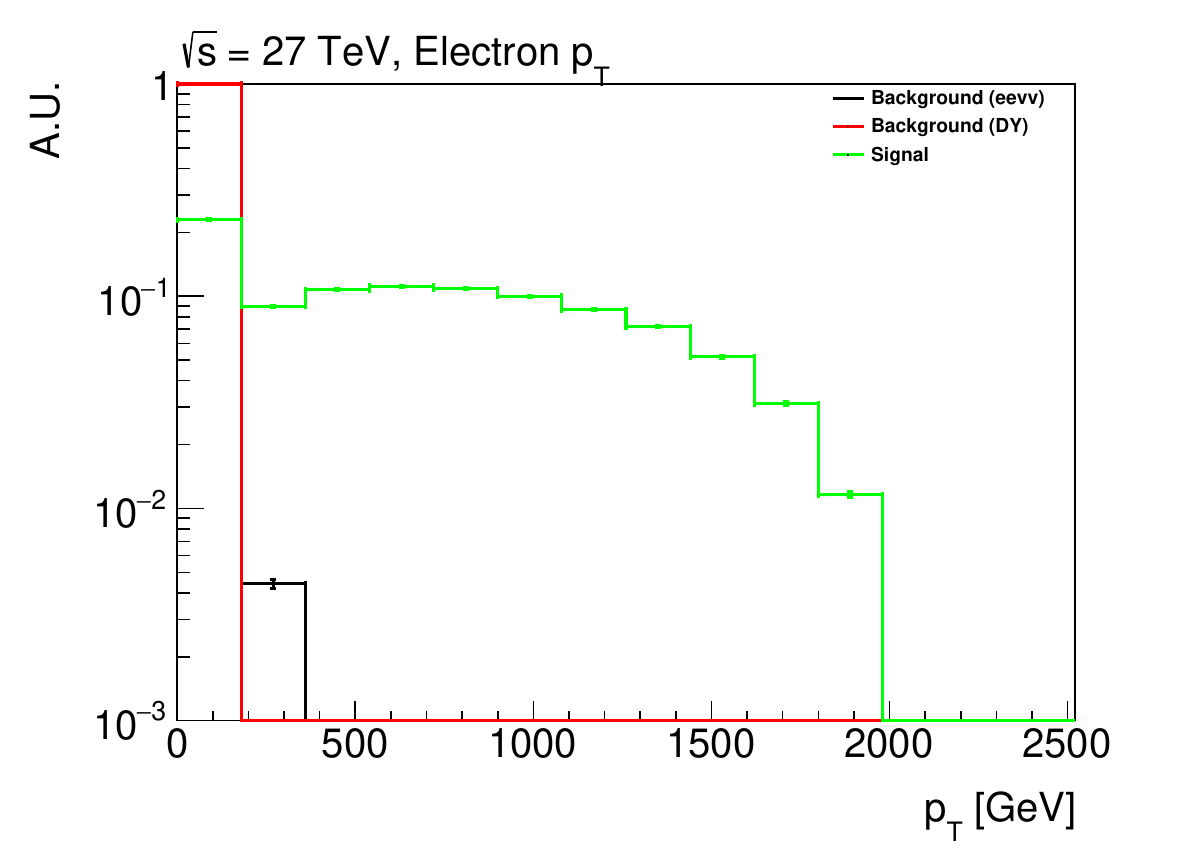}\qquad
    \includegraphics[scale=0.4]{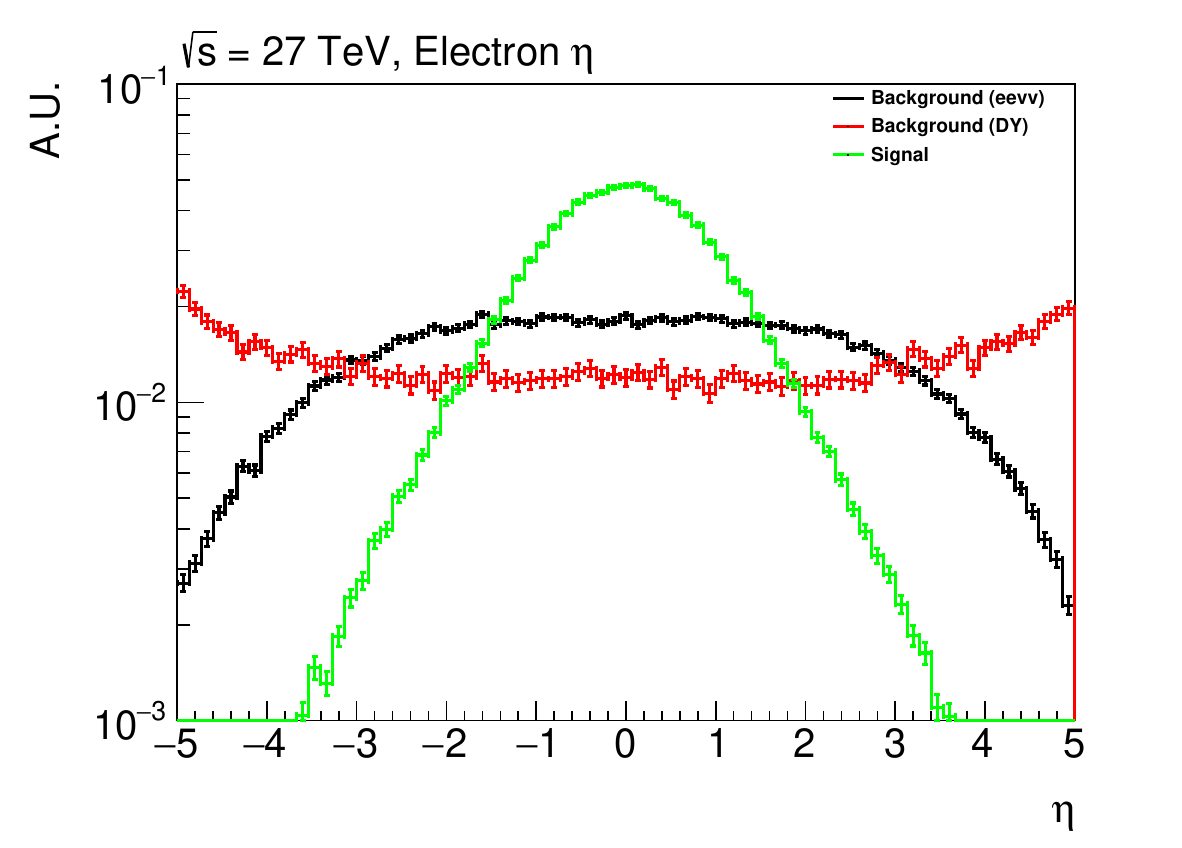}\qquad
    \includegraphics[scale=0.4]{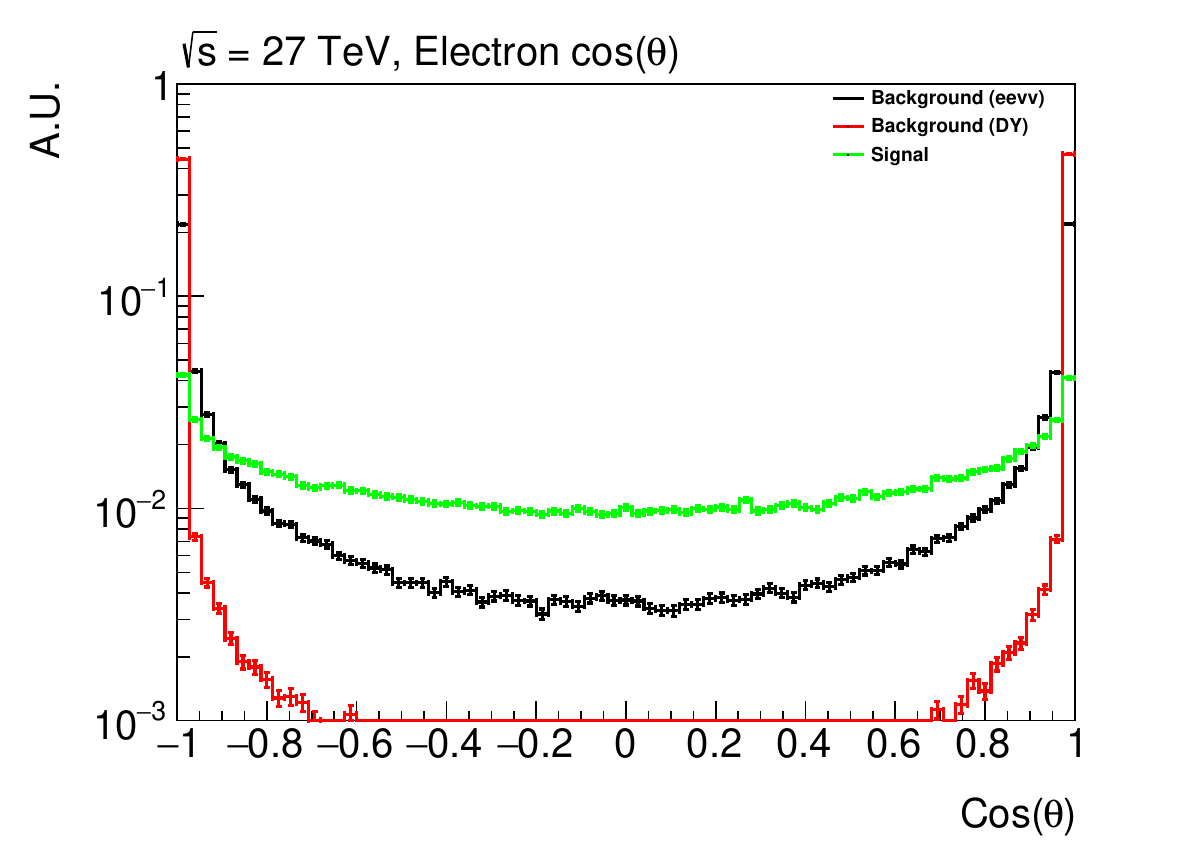}
    \caption{Kinematic distributions for signal and background for processes {\bf 2a}. We show the electron energy (top left), the electron transverse momentum (top right), the pseudorapidity (bottom left) and the angular separation between the outgoing electron and incoming proton beam (bottom right), before cuts. One can see that, in all plots, the signal (in green) rises above the background in some parameter region.}
\label{fig:hist2a}
\end{figure}
Fig. \ref{fig:hist2a} shows the energy, transverse momentum and angular distributions of final state electron for this case. We have normalised the events to unity, so as to pick the kinematic region advantageous to the signal events. We show the transverse momentum (top right plot) and the angular separation  between the electron and colliding proton beam in the bottom right figure ($\cos\theta$). The energy distribution is given in top left plot, and the pseudorapidity in the bottom left. As mentioned earlier, we are working with a centre-of-mass energy of $\sqrt{s}=27$ TeV, as expected as the HE-LHC. Study of kinematic distributions reveal that the energy and transverse momentum distributions are the best discriminant of signal against the background. Based on these kinematic distributions, we employ selection criteria of picking the central region in the $\cos\theta$ or pseudorapidity, and leave out events with energy less than 550 GeV, and transverse momentum less than $200$ GeV. The number of signal and background events expected, before and after the selection is given in Table~\ref{tab:selection2a} for an integrated luminosity of 30 fb$^{-1}$. Notice that the large cross section yields a large sensitivity even at such small luminosities.  However, one must keep in mind that this analyses is meant to be a feasibility study carried out at parton level. Inclusion of realistic detector effects and additional backgrounds that may arise from processes with different final states capable of mimicking the signal process will dilute this effect. 
\begin{table}
    \begin{center}
\caption{\label{tab:selection2a} Imposed cuts on both the signal and background and the number of signal and background events at 30 fb$^{-1}$ and $\sqrt{s} = 27$ TeV and 13 TeV for case {\bf 2a}, $pp\to (W_RW_R+W_RH_2^\pm+H_2^+H_2^-)\to 2e +$MET along with signal significance in each case.\\}
  \small   
   \begin{tabular}
   {l|l|c|c|c}
\hline\hline
&&&&\\
Case & Selection & No. of signal & Background events ($B$) & Significance \\
&  &events ($S$) & $pp\to ee\nu\nu$ & $s$ \\[2mm]
 \hline
 & &&&\\ 
{\bf 2a}&no cut: @ $\int{\cal L}=30$ fb $^{-1}$& 149&76020&0.5\\[2mm]\cline{2-5}
& &&&\\ 
$\sqrt{s}=$&$p_T>$ 200 GeV  &113&243&6.6\\[2mm]\cline{2-5}
27 TeV& &&&\\ 
&$p_T>$ 200 GeV,  &&&\\[1mm]
&$E>$ 550 GeV, $|\eta|<1.6, |\rm{cos}(\theta)| < 0.9$  &85&24&15.9\\[2mm]
\hline\hline
 & &&&\\ 
{\bf 2a}&no cut: @ $\int{\cal L}=300$ fb $^{-1}$& 110&318900&0.0006 \\[2mm]\cline{2-5}
&& &&\\ 
$\sqrt{s}=$& $p_T>$ 200 GeV  &80 &1020 &  2.3 \\[2mm]\cline{2-5}
13 TeV&&&&\\
&$p_T>$ 200 GeV,  &&&\\[1mm]
& $E>$ 550 GeV, $|\eta|<1.6, |\rm{cos}(\theta)| < 0.9$  &60 & 100 &  5.5 \\[2mm]
\hline\hline
\end{tabular}
\end{center}
\end{table}

The large signal sensitivity of this process at $\sqrt{s}=27$ TeV leads us to consider the possibilities of observing the signal at the LHC itself. We, therefore, consider the case at $\sqrt{s}=13$ TeV and report the expected number of events and signal significance at a slightly larger luminosity of 300 fb$^{-1}$ in the second half of the Table \ref{tab:selection2a}. The production cross section times branching fraction in this case is 0.37 fb, leading to 110 signal events before imposing any selection criteria. The corresponding background cross section is 1.063 pb, providing more than 0.3 million events at the luminosity of 300 fb$^{-1}$ considered. A significance of 5.5 $\sigma$ can be achieved with suitable kinematic selections. As noted above, this is expected to be diluted with realistic effects, which are not considered in this study.

\subsubsection{Case {\bf 2b}: $pp\to (W_RW_R+W_RH_2^\pm+H_2^-H_2^+)\to 2\tau +$ MET }

We revert back to the mass hierarchy of $m_{n_\tau} < m_{n_\mu}<m_{n_e}$, and consider the same parameters as we had considered in the cases {\bf 1a} and {\bf 1b}, except that for the exotic quark masses, which are taken be $m_{d'} = 1.3,~m_{s'}=1.4,m_{b'}=1.5$ TeV, heavier than the $W_R$ boson, with $M_{W_R}=1146$ GeV. The neutral dark scalars, $H_1^0$ and $A_1^0$ are heavier than the scotinos as well as the charged scalar $H_2^\pm$, leaving $n_\tau$ as the stable dark matter candidate particle. In this case, the dominant decays of $H_2^\pm$ as well as that of $W_R$ are to leptons and scotino. We shall consider the channel with lightest  scotino, $n_\tau$, along with $\tau$ lepton, as the simplest possibility. It will be hard to differentiate between the two boson productions ($W_R$ vs $H_2^+$), and therefore we shall analyse them together. 

\begin{figure}[htb!]
    \centering
    \includegraphics[scale=0.5]{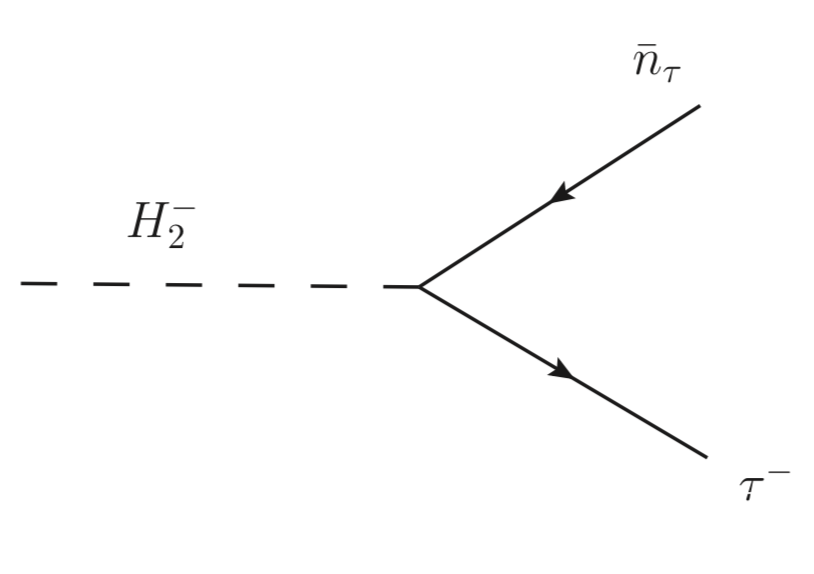}\qquad
    \includegraphics[scale=0.5]{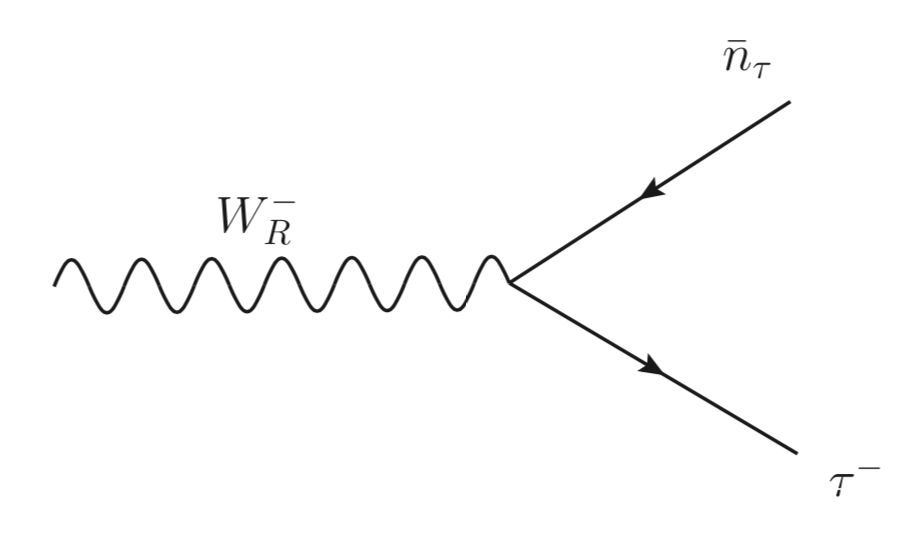}
    \caption{Decays of $H_2^-$ and $W_R$ for the process $pp\to (W_RW_R+W_RH_2^\pm+H_2^-H_2^+)\to 2\tau + $MET, case {\bf{2b}}.}
    \label{fig:process2b}
\end{figure}
 
As in the earlier cases, there are three possible productions, $W_RW_R$, $H_2^+H_2^-$ and the associated production $W_RH_2^\pm$, all leading to the final state is $2\tau+$MET. The cross sections for this process, along with the relevant branching ratios of $W_R$ and $H_2^\pm$ for the specific benchmark point considered are listed in Table \ref{tab:parameters2}. The SM background for this process is $pp \to 2\tau \nu\bar\nu$, with a cross section of $2.35$ pb.
\begin{figure}
    \centering
    \includegraphics[scale=0.42]{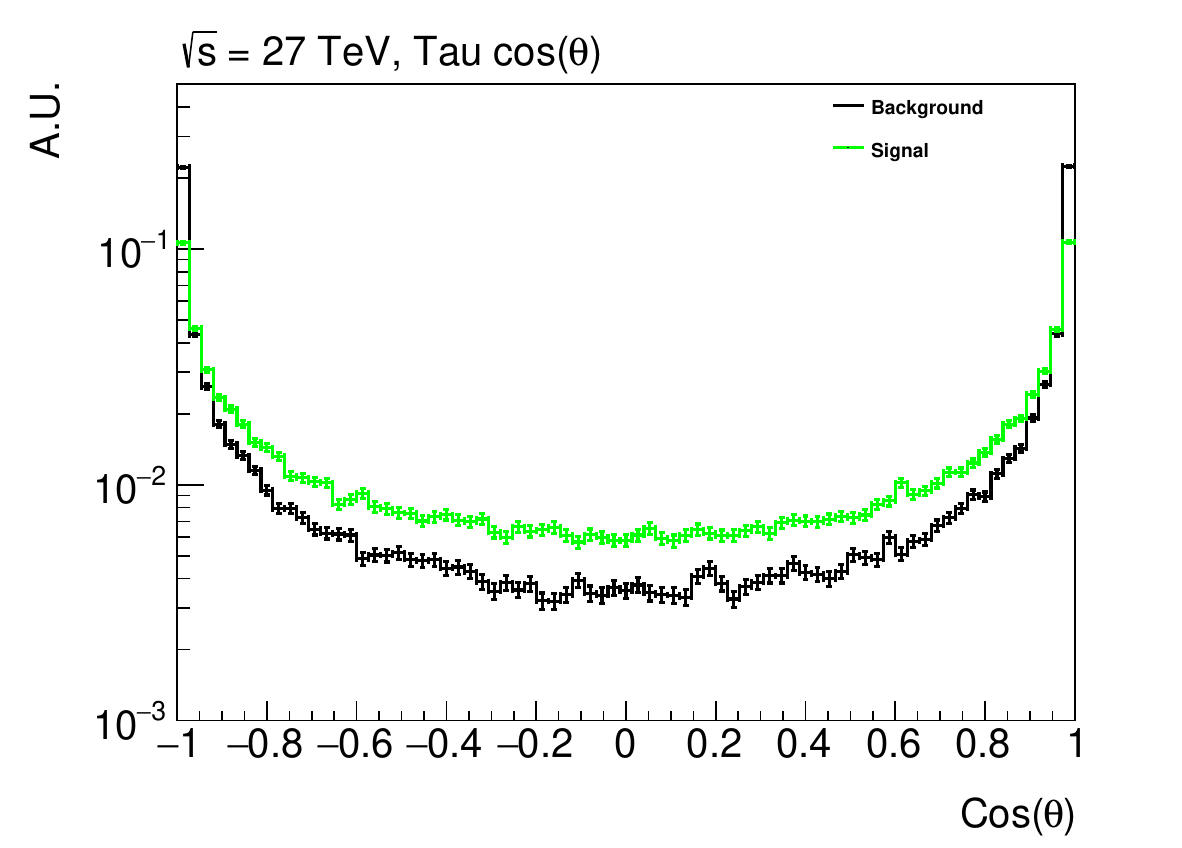} \qquad
    \includegraphics[scale=0.42]{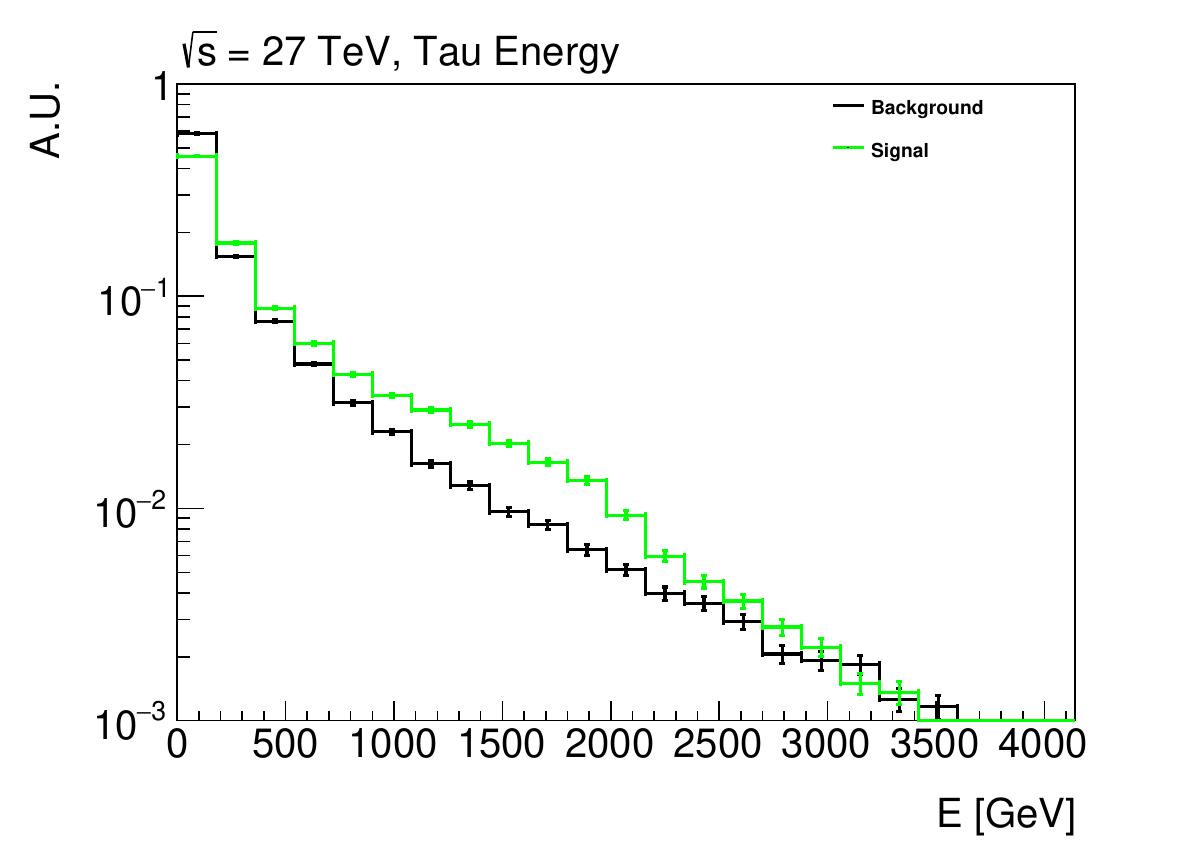} \qquad
    \includegraphics[scale=0.42]{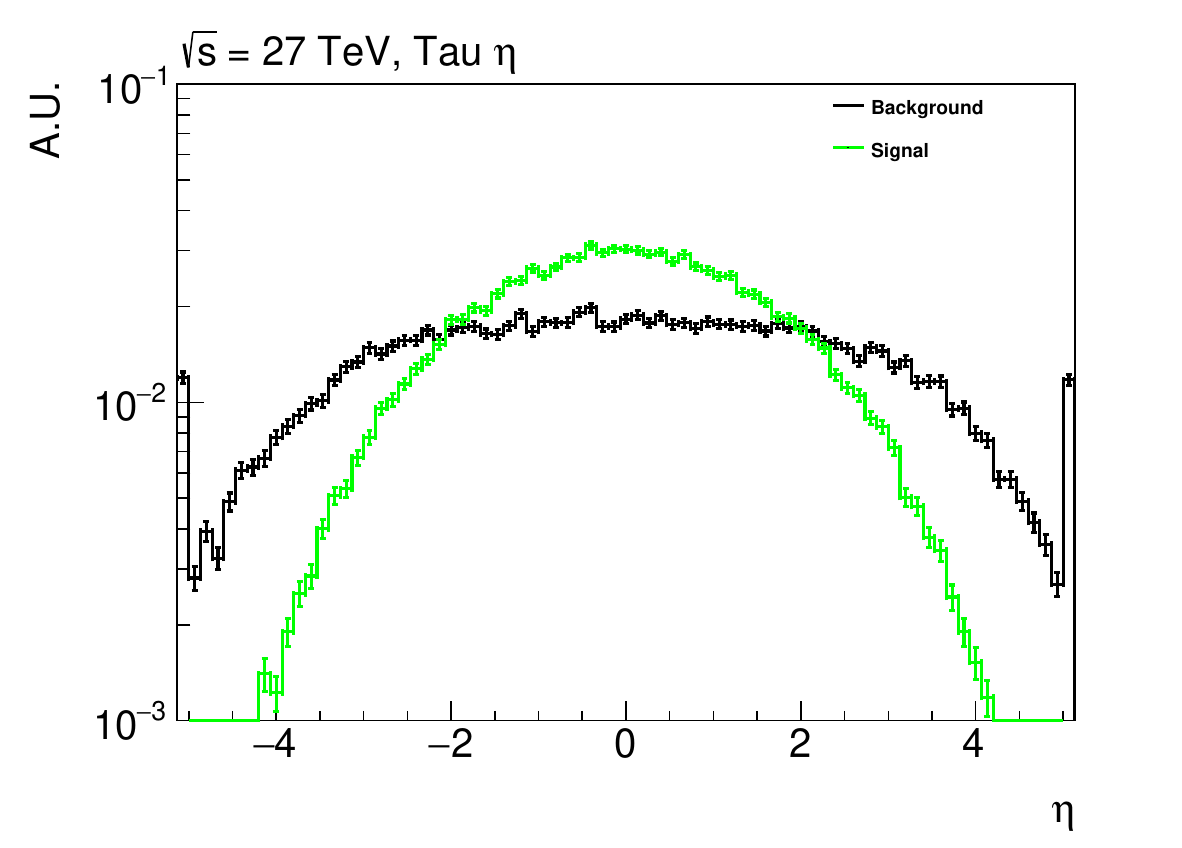}\qquad
    \includegraphics[scale=0.42]{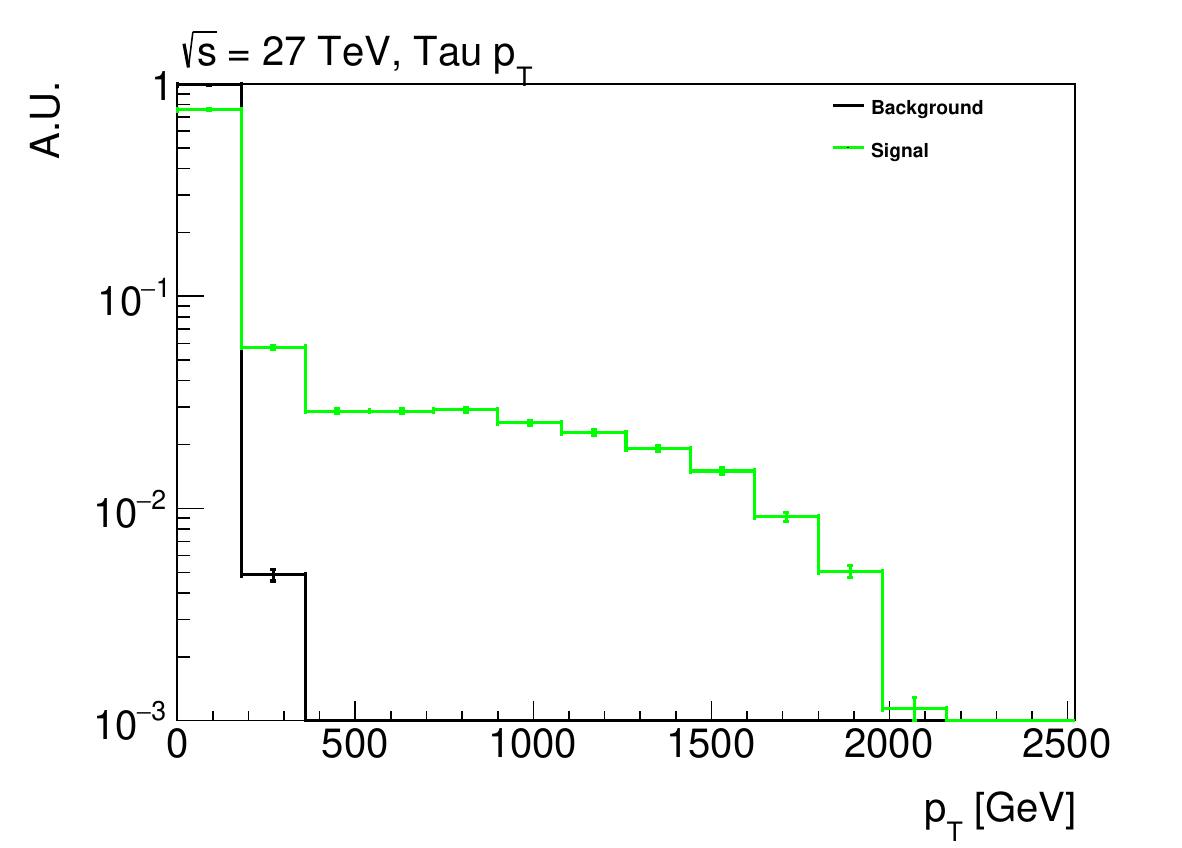}
    \caption{Signal and background for processes  {\bf 2b}. We show the angular separation  between the tau and incoming proton beam (top left), the tau energy (top right), the pseudorapidity (bottom left) and the tau transverse momentum (bottom right), before cuts. One can see that, in all plots, the signal (in green) rises above the background in some parameter region.}
    \label{fig:hist2b}
\end{figure}
Fig. \ref{fig:hist2b} shows the kinematic distributions of both the signal (green) and background (black) events at  $\sqrt{s}=27$ TeV, the variables used being the same as in case {\bf 1a} and {\bf 1b}. As expected, the $p_T$ of tau lepton peaks at small values for the background, whereas for the signal it is more or less a uniform distribution all the way up to and slightly above 2 TeV. We make use of this feature to employ the selection criteria of $p_T> 190$ GeV, which reduces the background considerably. This selection, however, keeps more than 77\% of the signal events. The number of signal and background events expected, before and after the selection, are given in Table~\ref{tab:selection2b} for an integrated luminosity of 30 fb$^{-1}$.
\begin{table}
    \begin{center}
\caption{\label{tab:selection2b} Imposed cuts on the backgrounds and signal and number of signal and background events at 30 fb$^{-1}$ for case {\bf 2b}, $pp\to (W_RW_R+W_RH_2^\pm+H_2^+H_2^-)\to 2\tau +$MET, along with signal significance in each case at $\sqrt{s}= 13$ TeV and 27 TeV.\\}
  \small   
   \begin{tabular}
   {l|l|c|c|c}
\hline\hline
Case & Selection & No. of signal & No. of background & Significance \\[1mm]
  & & ($S$) & ($B$) & ($s$)\\ [2mm]
 \hline
{\bf 2b} &no cut: @ $\int{\cal L}=30$ fb $^{-1}$& 3260&70500&11.2\\[2mm]\cline{2-5}
$\sqrt{s}=$&& &&\\  
27 TeV&$p_T>$190 GeV, $|\eta|<2.0$  &722&210&45.5\\[2mm]
\hline\hline
 && &&\\ 
{\bf 2b} & no cut: @ $\int{\cal L} = 30$ fb $^{-1}$ & 890 & 29420 & 4.7 \\[2mm]\cline{2-5}
 $\sqrt{s}=$&& &&\\
13 TeV& $p_T>$ 190 GeV, $|\eta|<1.6$ & 200 & 90 & 19.2 \\[2mm]
\hline\hline
\end{tabular}
\end{center}
\end{table}

As in the case of {\bf 2a}, the significance of this process at $\sqrt{s}=27$ TeV is very large. Below we show that the process is very promising even at 13 TeV LHC. The analyses results are summarised in Table \ref{tab:selection2b}, for 30 fb$^{-1}$ luminosity, showing a significance of 19$\sigma$. We may remind the reader that the feasibility study done here is at the parton level, without any detector effects and assuming 100\% identification and detection efficiencies. We are aware that these effects could considerably dilute the significance quoted here, especially in the case of {\bf 2b} with $\tau$ in the final state. 
\begin{table}[htb!]
    \begin{center}
\caption{\label{tab:CSMHp22a} The variation of cross-section with increasing  $M_{W_R}$ as well as $M_{H_2^+}$ masses for cases {\bf 2a} and {\bf 2b} for a fixed $g_R = 0.37$ at 27 TeV and 30 fb$^{-1}$ luminosity. Selection cuts applied are assumed to be  as in Tables \ref{tab:selection2a} and \ref{tab:selection2b}.  \\}
  \small   
   \begin{tabular}
   {c|c|c|c||c|c|c||c|c|c}
\hline\hline
&&&&\multicolumn{3}{c||}{Case {\bf 2a}}&\multicolumn{3}{c}{Case {\bf 2b}}\\ \cline{5-10}
&&&&&&&&&\\
  $ |\mu_3| $ (GeV) & $v^\prime$ (TeV) & $M_{W_R}$ (TeV) & $M_{H_2^+}$ (GeV) & $M_{H_1^0} =M_{A_1^0} $ & $\sigma\times BR$ (fb) & $s$ after cut & $M_{H_1^0} =M_{A_1^0}$ & $\sigma\times BR$ (fb) & $s$ after cut\\
  &&&& (GeV)&&& (GeV) &&
  \\[2mm]
 \hline
 & &&&&&&&\\ 
 450 & 6.2 & 1.15 & 199 & 165 & 3.6 & 11.5& 316 & 94&39.3 \\[2mm]\hline
 &&&&&&&&\\
 850 & 6.2 & 1.15 & 273 & 250 & 5.3 & 17.0& 368 & 47&19.7 \\[2mm] \hline
 &&&&&&&&\\
 400 & 6.5 & 1.203 & 192 & 157 & 2.9 &  9.3& 312 & 99&41.4 \\[2mm] \hline
 &&&&&&&&\\
 700 & 6.5 & 1.203 & 254 & 228 & 3.4 & 10.8& 353 & 49&20.5 \\[2mm] \hline
 &&&&&&&&\\
 400 & 7.5 & 1.39 & 206 & 174 & 1.7 & 5.4& 321 & 72&30.1 \\[2mm]\hline
 &&&&&&&&\\
 400 & 8.5 & 1.57 & 219 & 190 & 0.89 & 2.9 & 330 & 54&22.7 \\[2mm]\hline
&&&&&&&&\\
 800 & 8.5 & 1.57 & 310 & 290 &  -- & -- & 396 &17 &7.1 \\[2mm] \cline{1-10}
&&&&&&&&\\
 400 & 10 & 1.85 & 238 & 211 & 0.46 & 1.5 & 342 & 38 & 15.8\\[2mm]\hline
&&&&&&&&\\
900 & 10 & 1.85& 357& 339 & --&--& 433 & 8.7&3.6\\[2mm]
\hline\hline
\end{tabular}
\end{center}
\end{table}

Further, all the above analyses are done, as before, with a fixed  $M_{W_R}=1146$ GeV and for charged Higgs mass of $M_{H_2^+}=187$ GeV. Motivated by large achievable significance, we see possibilities of detecting a signal with heavier $W_R$ and $H_2^+$. For this, we vary $M_{W_R}$ and $M_{H_2^+}$ by changing $v'$ and $\mu_3$, respectively, keeping all other parameters the same as in the previous analysis.  For specific choices of $M_{W_R}$ and $M_{H_2^+}$, the results are presented in Table \ref{tab:CSMHp22a} for the cases of {\bf 2a} and {\bf 2b} at $\sqrt{s}=$ 27 TeV.  The effective cross section is more sensitive to $M_{H_2^+}$ than to $M_{W_R}$, as the charged Higgs boson is allowed to be much lighter. Even with luminosity as small as 30 fb$^{-1}$, it is possible to probe $M_{W_R}$ up to about 2 TeV along with $M_{H_2^+}$ between 200 and 300 GeV. In this table we have left the values of $\sigma \times$ BR and $s$ after cuts for two BPs blank for process {\bf 2a} as these BPs are not compatible with the scenario considered in the process i.e. $M_{H_2^+} < m_{n_e}$.

\subsection{Case {\bf 3} $M_{W_R}<~(m_{d^\prime}, m_{s^\prime}, m_{b^\prime}$,~~$m_{n_e},m_{n_\mu},m_{n_\tau})$}
\label{subsec:3}                                                        

So far, we have analysed two different mass hierarchies. In all cases the scotinos were lighter than the $W_R$, and in case {\bf 1} the exotic quarks $(d')$ were taken to be lighter than $W_R$ while in case {\bf 2} this mass hierarchy between $W_R$ and exotic quarks was reversed. The major decay channels of $W_R$ in each case were then to either the exotic quarks (case {\bf 1}) or to scotinos (case {\bf 2}). We  now explore the parameter regions with $W_R$ lighter than both the scotinos and the exotic quarks, disallowing two body decays of $W_R$ into either of these particles. The additional Higgs bosons are still lighter than $W_R$, favouring its decay into these scalars along with either the standard gauge bosons ($W_L,~Z$) of the SM Higgs boson $h$. We probe here this scenario through the process
\begin{equation}
    pp \rightarrow (W_R^+ W_R^- + W_R^+ H_2^- + H_2^+ H_2^-) \rightarrow 2j + e^- + ~\rm{MET}
\end{equation}
\begin{table}[htbp]
\caption{\label{tab:parameters3}Parameter values for process {\bf 3}. We list, in the first row, the relevant parameters in the scalar potential Lagrangian, in the second row, the masses for the exotic quarks and the scotinos, in the following row the the  masses for bosons $W_R$, $Z^\prime$, $H_2^\pm$ and the the dark matter particles $H_1^0$, $A_1^0$. Below we list the relevant branching ratios for $W_R$  yielding the final states, and the production cross sections for process {\bf 3}.  We have assumed in both cases the minimum $p_T$ for both jets and leptons to be 10 GeV.}
  \begin{center}
 \small
 \begin{tabular*}{1.0\textwidth}{@{\extracolsep{\fill}}ccccccc}
 \hline\hline
	$\lambda_2$ & $\lambda_3$ &$\alpha_1=\alpha_2=\alpha_3$ &$\mu_3$ &$\tan \beta$ &$g_R$ &$v^\prime$ 
	\\
	\hline
	-0.1 & 1.6 & 0.01 &-400 GeV & 50 & 0.37 & 6.2 TeV \\
  \hline\hline   
 $ m_{d^\prime}$ & $m_{s^\prime}$ & $m_{b^\prime}$ & $m_{n_e} $ & $m_{n_\mu}$ & $m_{n_\tau}$
\\ 
	 \hline
 1.3 TeV & 1.4 TeV  & 1.5 TeV &   1401 GeV & 1351 GeV	& 1301 GeV \\ 
\hline \hline
$M_{W_R}$ & $M_{Z^\prime}$ &  $M_{H_2^-}$ &$M_{H_1^0}=M_{A_1^0}$ &  &
 \\	 
\hline
 1.146 TeV & 4.55 TeV &187 GeV&151 GeV & \ &  \\	 
  \hline \hline
      \end{tabular*}
 \begin{tabular*}{1.0\textwidth}{@{\extracolsep{\fill}}ccccc}
    BR($W_R \rightarrow W_L~H_1^0/A_1^0$)  & BR($W_R \rightarrow ZH_2^+$)  & BR($W_R \rightarrow hH_2^+$) 
   &BR($W_R \rightarrow W_LW_LH_2^+$&BR($W_R \rightarrow \bar tbH_2^+$))
  \\ 
  \hline
 49.36 \% &  24.29\%   &  22.06\% &1.31\%&1.49\%\\   \hline 
 &&&&\\
 BR($H_2^+\to jj~H_1^0/A_1^0$)& BR($H_2^+\to e\nu~H_1^0/A_1^0$)
  &~~~signal $\sigma\times$BR($e^-~jj~MET$)&~~~SM~$\sigma$ ($pp \rightarrow 2j + e^- + \bar{\nu}$)  \\  
  \hline
 67.04\%&11.18\%& 34.1~fb&$6.4\times 10^7$ fb  \\
  \hline \hline
    \end{tabular*}
\end{center}
 \end{table}

With the above benchmark point, decay branching fractions of $W_R$ and $H_2^+$ are given in Table \ref{tab:parameters3}. Adding all channels, $pp \rightarrow (W_R^+ W_R^- + W_R^+ H_2^- + H_2^+ H_2^-) \rightarrow 2j + e^- + \rm{MET}$ has a cross section of 34.1 fb. The SM background has a cross section of $\sigma( pp \rightarrow 2j + e^- + \bar{\nu})=  64011.58~ \rm{pb}$. This process proceeds according to the Feynman diagrams of Figure \ref{fig:feyn3b}.
\begin{figure}[htb!]
    \centering
    \includegraphics[scale=0.3]{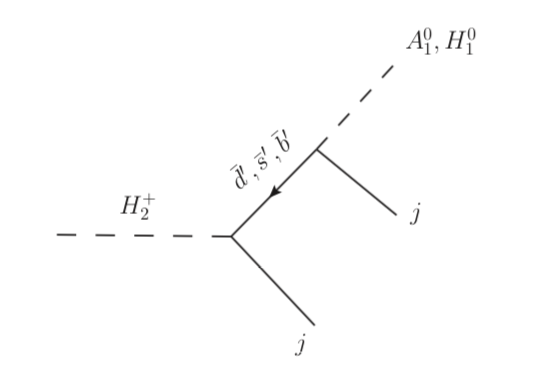}\qquad 
    \includegraphics[scale=0.3]{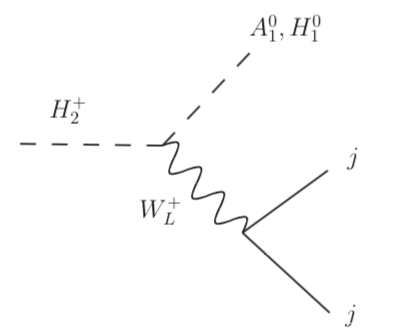}\qquad
    \includegraphics[scale=0.3]{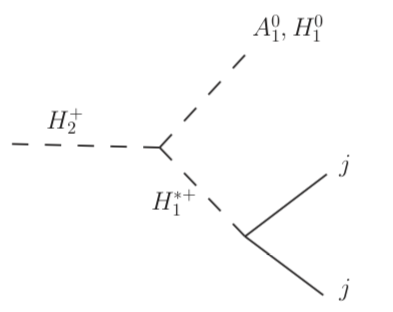}\qquad
    \includegraphics[scale=0.3]{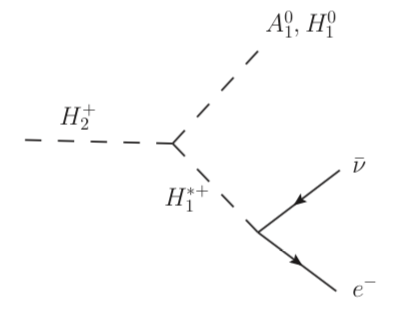}\qquad
    \includegraphics[scale=0.3]{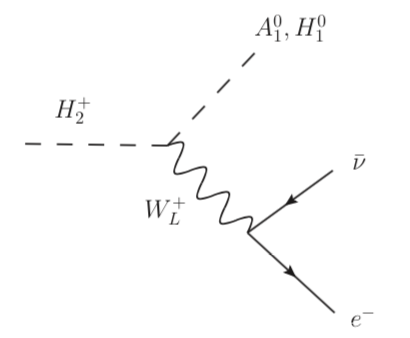}\qquad
    \includegraphics[scale=0.3]{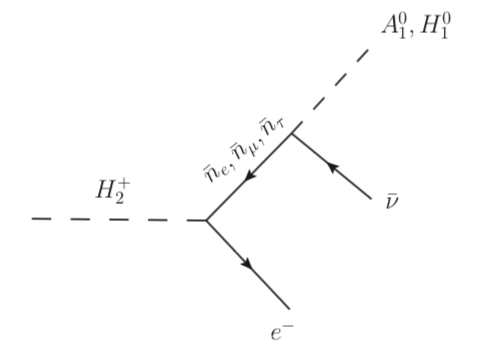}
    \caption{Feynman diagrams for $H_2^+$ decay for process {\bf 3}, $pp\to (W_RW_R/W_RH_2^\pm/H_2^+H_2^-)\to 2j+e^-+$MET.}
    \label{fig:feyn3b}
\end{figure}

\begin{figure}[htb!]
    \centering
    \includegraphics[scale=0.25]{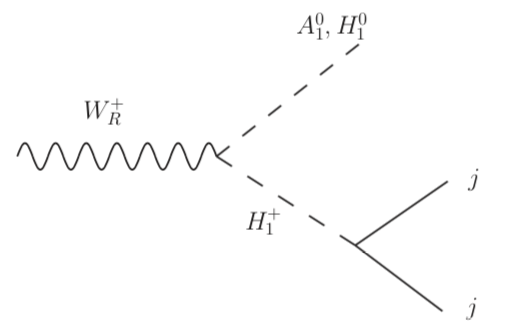}\qquad 
    \includegraphics[scale=0.25]{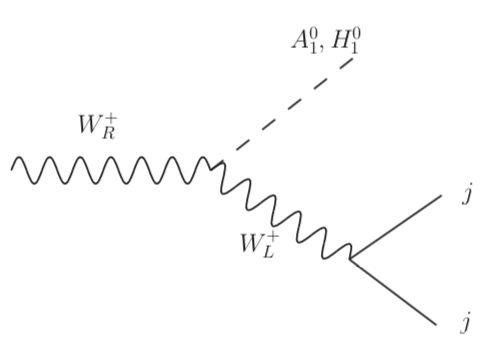}\qquad
    \includegraphics[scale=0.25]{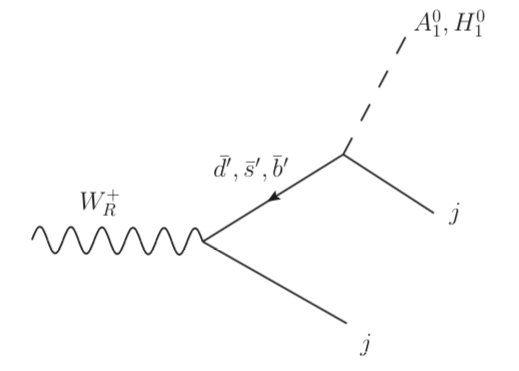}\qquad
    \includegraphics[scale=0.25]{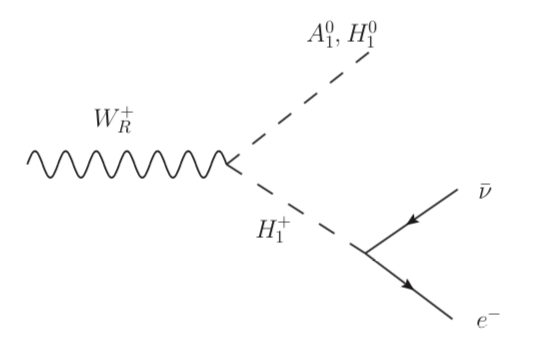}\qquad
    \includegraphics[scale=0.25]{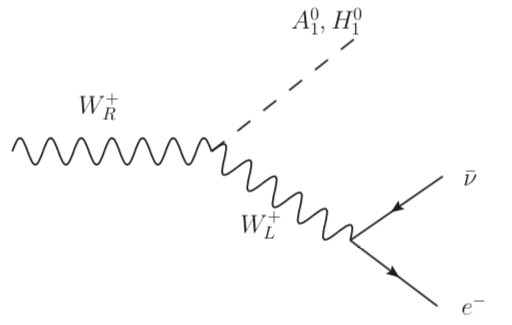}\qquad
    \includegraphics[scale=0.25]{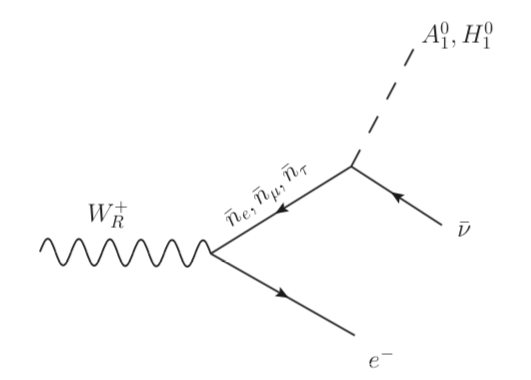}\qquad
    \includegraphics[scale=0.25]{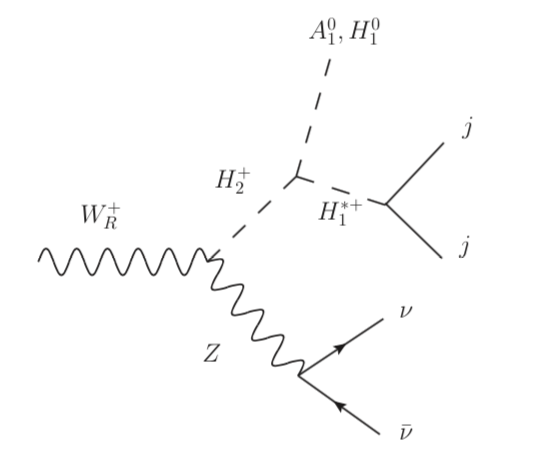}\qquad 
    \includegraphics[scale=0.25]{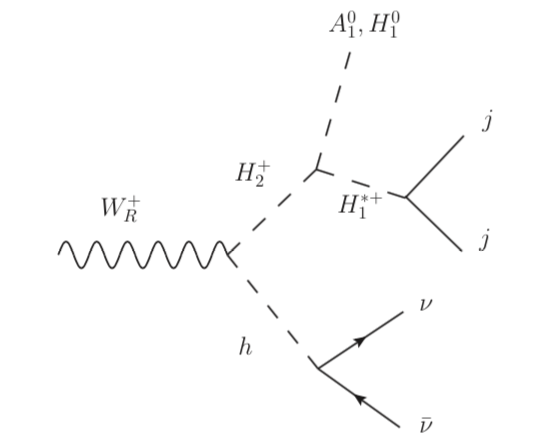}\qquad
    \includegraphics[scale=0.25]{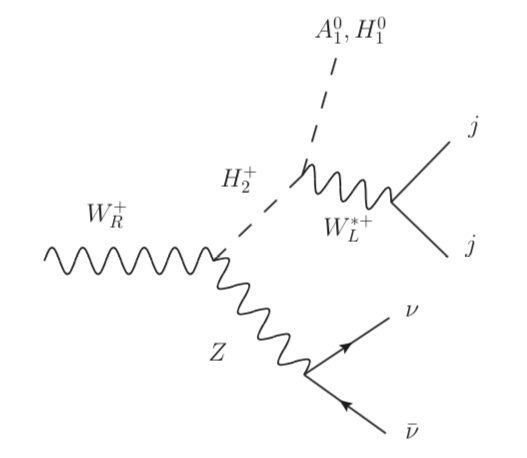}\qquad
    \includegraphics[scale=0.25]{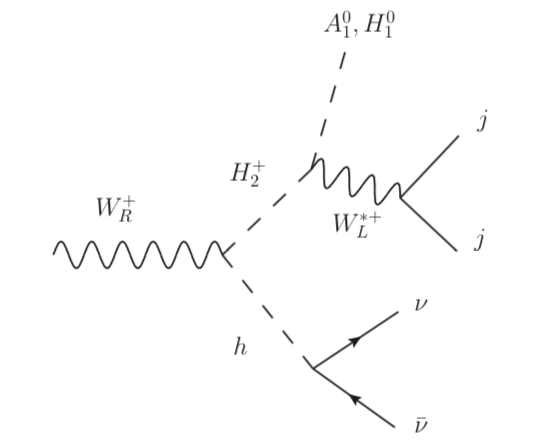}\qquad
    \includegraphics[scale=0.25]{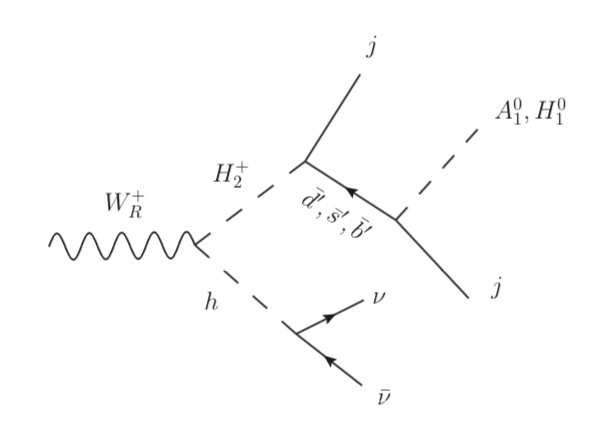}\qquad
    \includegraphics[scale=0.25]{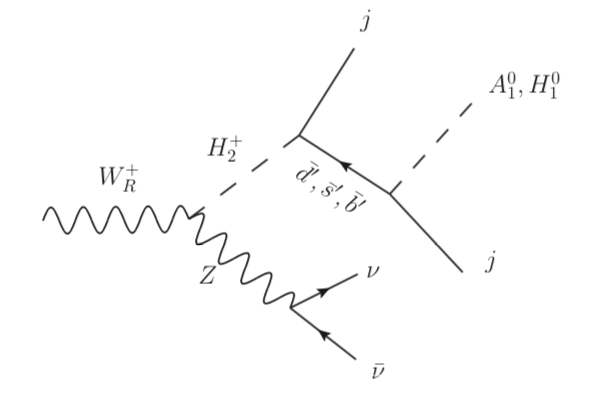}
    \caption{Feynman diagrams for $W_R^+$ decay for process {\bf 3}, $pp\to (W_RW_R/W_RH_2^\pm/H_2^+H_2^-)\to 2j+e^-+$MET.}
    \label{fig:feyn3b}
\end{figure}

With such large background, kinematic selections are necessary to look for the signal events. 
\begin{figure}[htb!]
    \centering
    \includegraphics[scale=0.42]{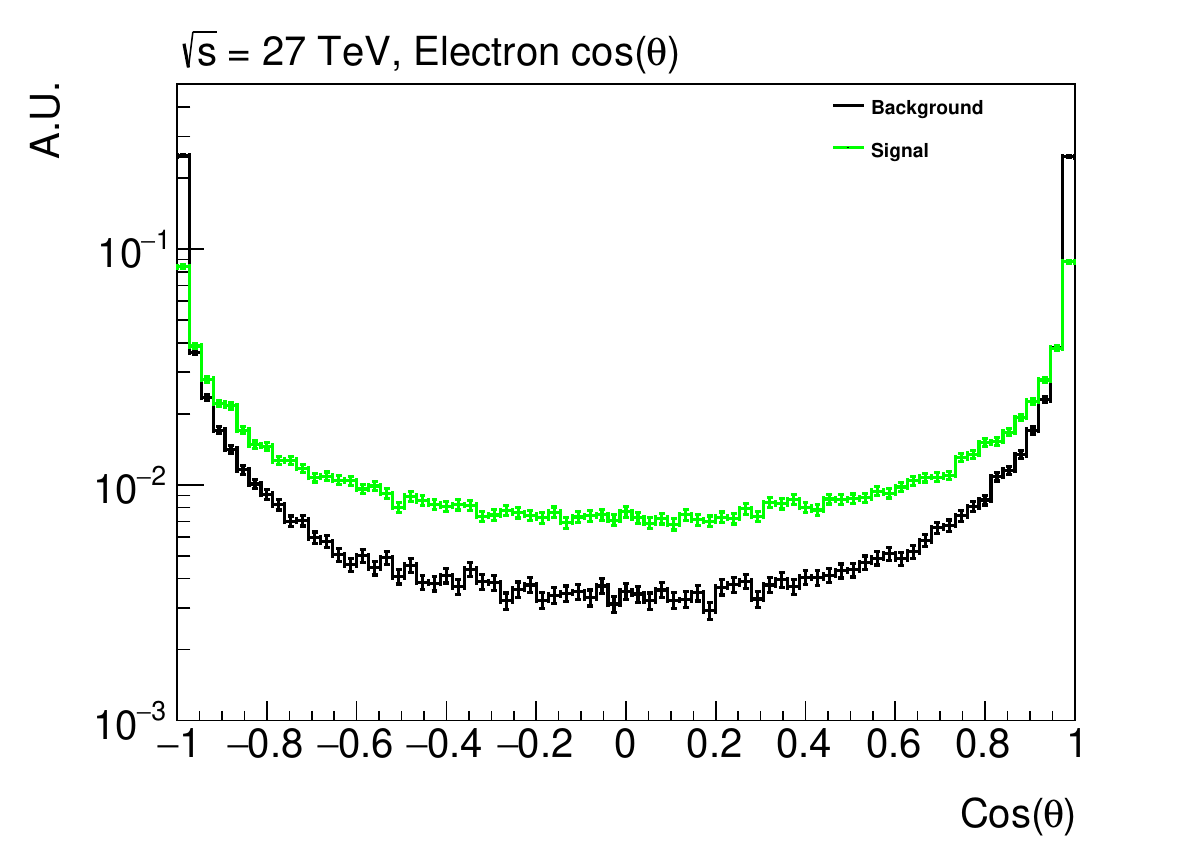}\qquad
    \includegraphics[scale=0.42]{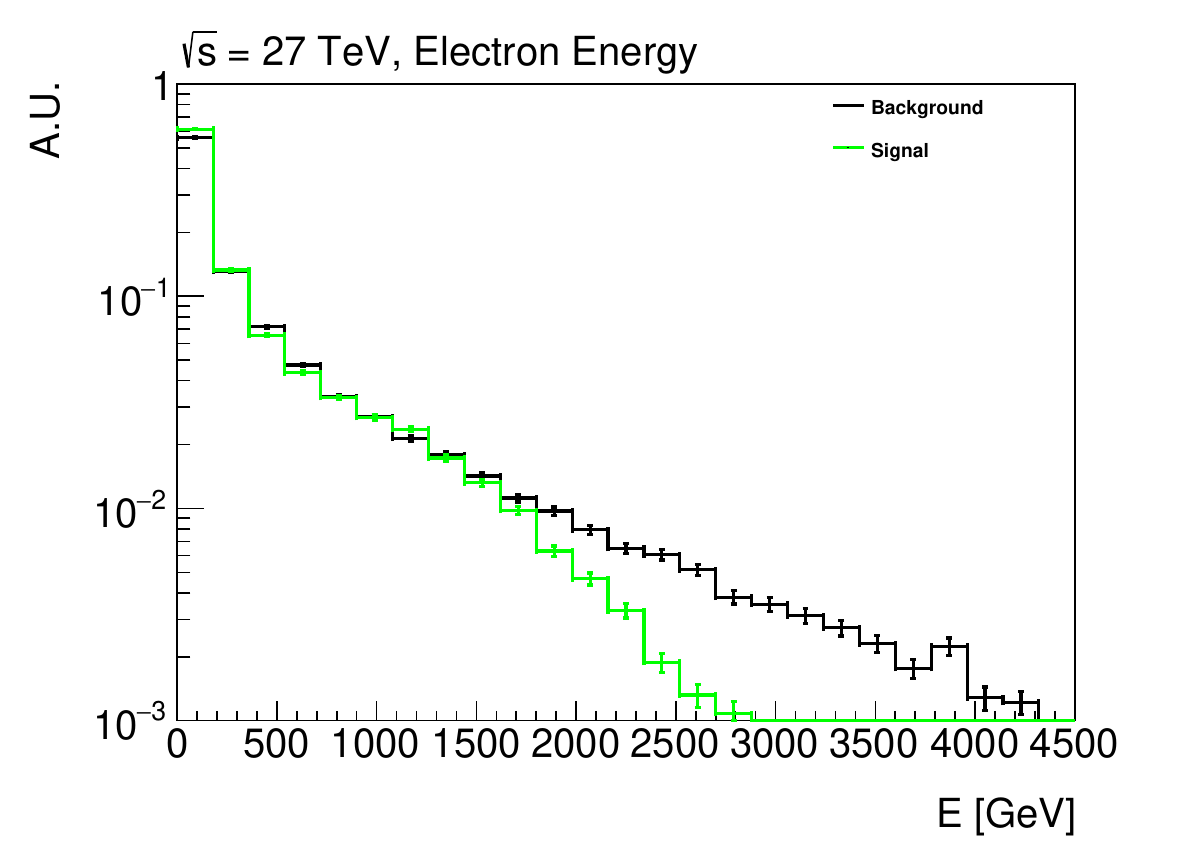}\qquad
    \includegraphics[scale=0.42]{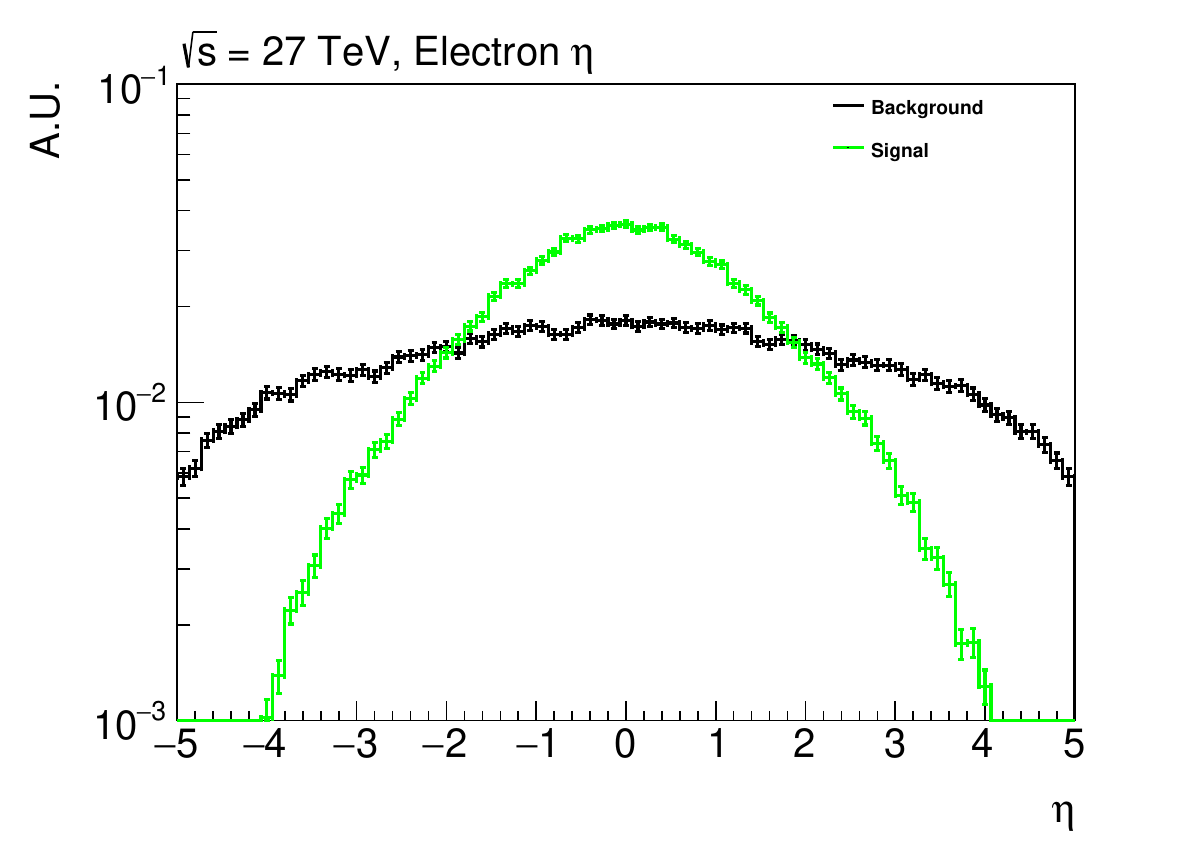}\qquad
    \includegraphics[scale=0.42]{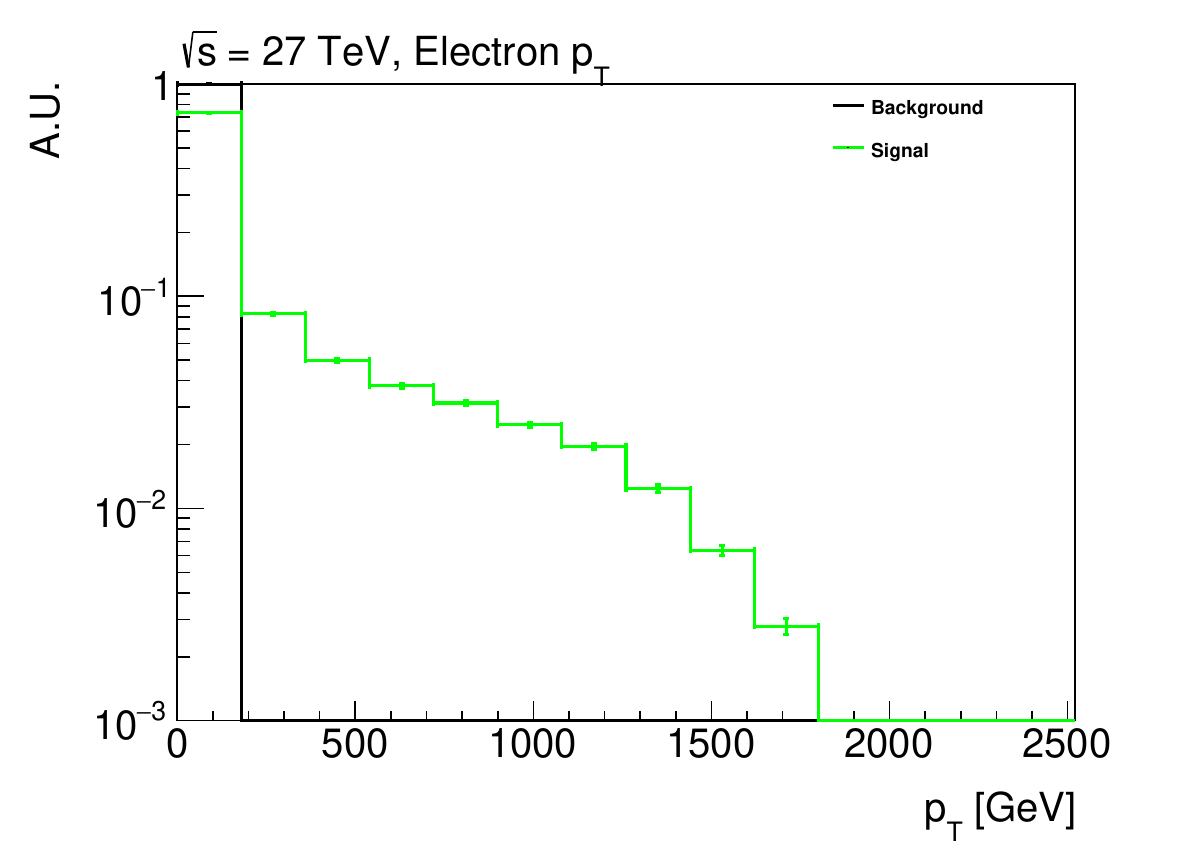}
    \caption{Signal  plots for process {\bf 3} for $\sqrt{s}=27$ TeV. We show the angular separation between the final state electron and incoming proton beam (top left), the electron transverse energy (top right), the pseudorapidity (bottom left) and the electron transverse momentum (bottom right), before cuts. The signal is shown in blue as there is no such SM background arising in this scenario producing same final state particles i.e., $pp \rightarrow 2j + e^- + \rm{MET}$.}
    \label{fig:hist3b}
\end{figure}
\begin{table}  
\begin{center}
\caption{\label{tab:selection3b} Imposed cuts on the signal and the background kinematic distributions and number of signal and background events 
at 300 fb$^{-1}$ for case {\bf 3}, $pp\to (W_RW_R+W_RH_2^\pm+H_2^+H_2^-)\to 2j + e^- +$MET, along with signal significance in each case. \\}
\begin{tabular}
{l|l|c|c|c}
\hline
\hline
 && &&\\ 
Case & Selection & No. of signal ($S$)
& No. of background ($B$) & Significance ($s$)\\[2mm]
 \hline
  && &&\\ 
{\bf 3}&no cut: @ $\int{\cal L}=3000$ fb $^{-1}$& 
102300&1.92 $\times 10^{11}$&0.09 \\[2mm]\cline{2-5}
&& &&\\  
&$p_T>$ 190 GeV, $|\eta|<2.0$  &
26119&6.14 $\times 10^7$&3.3 \\[2mm]
\hline\hline
\end{tabular}
\end{center}
\end{table}
The kinematic plots in Fig. \ref{fig:hist3b} guide us with this selection. Imposing an event selection of $p_T(e) > 190$ GeV and electron pseudo-rapidity, $|\eta|<2$ reduces the background to 0.3 per mill, while keeping more than 25\% of the signal events. This leads to a significance of about 3.3 at 3 ab$^{-1}$ luminosity, indicating that this signal is rather difficult to probe.

\section{Conclusion}
\label{sec:conclusion}
In this work we analyzed the possibility of discovering signals indicating the presence of a charged gauge vector boson at the HE-LHC and HL-LHC. A charged gauge vector boson is indicative of a BSM scenario  which includes a symmetry higher than $U(1)$, the simplest of which would be an extra $SU(2)$ gauge group. The most commonly known and frequently analyzed group is $SU(2)_R$ of the left-right symmetric model, LRSM, where right-handed fermions mirror left-handed ones from the SM--that is they belong to $SU(2)_R$ doublets. The disadvantage of such models is that the charged gauge boson $W^\prime$ (an admixture of the $SU(2)_R$ $W_R$ and the SM $W_L$ bosons) must be heavy to avoid flavour-changing neutral currents in the kaon and $B$-meson systems. Recent analyses indicate  that to observe this boson would require a 100 TeV collider \cite{Chauhan:2018uuy}.

However, in an alternative left right model, ALRM, emerging from the breaking of $E_6$, the corresponding gauge group would be $SU(2)^\prime_{R}$. Here,  instead of partnering with the right-handed down-type quark, the right-handed up-quark joins with a new exotic coloured fermion, $d^\prime_R$ (of the same charge as $d_R$), to form a doublet under the $SU(2)^\prime_{R}$, and similarly, the right-handed charged lepton partners with a new neutral fermion ($n_R$) to form a doublet under the same $SU(2)^\prime_{R}$. The right-handed down-type quark, $d_R$ and the right-handed neutrino, $\nu_R$, along with the left-handed degrees of freedom of the newly introduced fermions, $d_L^\prime$ and $n_L$, remain singlets under both $SU(2)_L$ and $SU(2)^\prime_{R}$. 

This model presents several advantages over the better known left right symmetric model. It allows for the introduction of a generalized lepton number and identifying a natural $R$-parity without supersymmetry. It provides  new neutral fermions and neutral Higgs which can serve as dark matter candidates which, due to the $R$-parity symmetry, are stabile. Unlike in LRSM, the symmetry of the model disallows the $W_R$ gauge boson to mix with the SM $W_L$ boson. It avoids flavour-changing neutral currents, allowing charged Higgs bosons and the $W_R$ boson to be lighter. 

This is the advantage we explore here in analysing the imprint of the $W_R$ boson at colliders. We perform a comprehensive analysis of the parameter space of the ALRM,  
for the two cases of the relative masses of the exotic quarks and leptons with respect to the $W_R$ bosons. The case where the exotic quarks are lighter, they enter the decay of the $W_R$ as intermediate states, while in the second case, being heavier, they do not.   
The first possibility does not appear to produce visible signals at 27 TeV, while the latter does. For the second case, where the $W_R$ bosons are lighter than the exotic quarks, the former can decay into purely leptonic states. We considered two different situations in this case, one with scotinos lighter than the scalars, and  another one with the scalars lighter. The degenerate dark neutral scalar and pseudoscalar bosons ($H_1^0$ and $A_1^0$) are the stable dark matter candidates in the latter situation, while the lightest scotino takes this role in the former case. We considered two different signals in  these cases; $e^+e^- +$ MET in the first case, and $\tau^+\tau^-+$ MET in the second case. Our analysis shows that both processes have large significance over the expected SM background. Studying a different benchmark points in each scenario, and varying $W_R$ and scalar masses, a signal significance larger than $3\sigma$ seems to be achievable for  $M_{W_R}\sim 1.9$ TeV and  $M_{H_2^\pm}$ around 350 GeV, even at a low luminosity of 30 fb$^{-1}$. In the third case we considered $M_{W_R}$ smaller than both $m_{d'}$ and $m_n$. In this case, $W_R$ decays through $H_2^\pm$ and $W_L$, and we chose a final state of $2j+e^-+$ MET for our analysis.  Despite the large background here, we shown that it is possible to reach more than $3\sigma$ significance with 3 ab$^{-1}$ luminosity at $\sqrt{s}=27$ TeV.

Thus, contrary to the LRSM $W_R$ gauge boson, which must be heavy, this analysis indicates a promising signature for the ALRM which would distinguish it from the LRSM, and an encouraging signature for the presence of charged gauge bosons, indicative of a higher gauge symmetry.

\vskip1cm
{\large \bf Acknowledgement}\\[5mm]
\label{acknowledgement}
We are grateful to Prof. Ernest Ma for suggesting this project. The work of MF has been partly supported by NSERC through grant number SAP105354. CM acknowledges the Royal Society, UK for the support through the Newton International Fellowship with grant number NIF$\backslash$R1$\backslash$221737. SS is partially funded for this work under the US Department of Energy contract DE-SC0011095. The work of PP is supported by SERB, DST, India through grant CRG/2022/002670.

\bibliographystyle{apsrev4-2}
\bibliography{ALRMWR}

\end{document}